\documentclass[aip,pof,12pt]{revtex4-2}

\usepackage{graphicx}
\usepackage{epstopdf, epsfig}
\usepackage{bm}
\usepackage{natbib}
\usepackage{ar}
\usepackage{amsmath}
\usepackage{amssymb}
\usepackage{rotating}
\usepackage{afterpage}

\newcommand{\vect}[1]{\textbf{\textit{#1}}} 
\newcommand{\vort}{\boldsymbol{\omega}} 
\newcommand{\uv}[1]{\hat{\boldsymbol{#1}}} 
\newcommand{\der}[2]{\frac{\mathrm{d}#1}{\mathrm{d}#2}} 
\newcommand{\pder}[2]{\frac{\partial#1}{\partial#2}} 
\newcommand{\del}{\boldsymbol{\nabla}} 
\newcommand{\grad}[1]{\del #1} 
\newcommand{\curl}[1]{\del\times\boldsymbol{#1}} 
\newcommand{\md}[1]{\mathrm{d}#1} 
\newcommand{\mdi}[1]{\hspace{2pt}\mathrm{d}#1} 
\newcommand{\cc}[1]{\overline{#1}} 
\newcommand{\e}[1]{\mathrm{e}^{#1}} 

\newcommand{\lrp}[1]{\left(#1\right)} 
\newcommand{\bp}[1]{\big(#1\big)} 
\newcommand{\lrc}[1]{\left\{#1\right\}} 
\newcommand{\lrb}[1]{\left[#1\right]} 
\newcommand{\bb}[1]{\big[#1\big]} 

\usepackage{hyperref}
\hypersetup{
    colorlinks=true,
    linkcolor=blue,
    urlcolor=blue,
    citecolor=blue,
}

\begin{document}


\title{A theoretical model for the separated flow around an accelerating flat plate using time-dependent self similarity} 



\author{A. C. DeVoria}
\affiliation{Department of Mechanical \& Aerospace Engineering, University of Florida, Gainesville, FL 32611, USA}

\author{K. Mohseni}
\email[]{mohseni@ufl.edu}
\affiliation{Department of Mechanical \& Aerospace Engineering, University of Florida, Gainesville, FL 32611, USA}
\affiliation{Department of Electrical \& Computer Engineering, University of Florida, Gainesville, FL 32611, USA}



\begin{abstract}
We present a model appropriate to the initial motion (2-3 chords of travel) of a flat-plate airfoil accelerating in an inviscid fluid. The separated flow structures are represented as vortex sheets in the conventional manner and similarity expansions locally applicable to the leading and trailing edges of the plate are developed. The topological character of vortex sheets is maintained rather than resorting to point vortex discretizations. Beyond this, there are two theoretical novelties to our approach as compared to previous studies. First, an expansion is applied to the attached outer flow rather than the vortex sheet circulations and positions. This allows the asymmetric effect of the sweeping component of the free-stream flow parallel to the plate to be built-in to the same governing equation as the singular-order flow. Second, we develop a time-dependent self similarity procedure that allows the modeling of more complex evolution of the flow structures. This is accomplished through an implicit time variation of the similarity variables. As a collective result, the predicted vortex dynamics and forces on the plate compare favorably to Navier-Stokes simulations. Lastly, the model is utilized to provide some new intuition about the separated flow at the leading edge.
\end{abstract}

\maketitle 

\section{Introduction}\label{sec:intro}
The two-dimensional flow around a flat-plate airfoil is a canonical problem of classical aerodynamics and has been extensively studied by theorists, experimentalists, and computationalists alike. Accordingly, it continues to serve as a benchmark for developing new, low-order inviscid modeling techniques that tackle physical problems with increasing complexity. In particular, a myriad of methods have been proposed for the vortex shedding from both the leading and trailing edges and the corresponding unsteady forces exerted on the plate, for example Refs.~[\onlinecite{KatzJ:81a,GrahamJMR:83a,JonesMA:03a,LlewellynSmith:09a,Mohseni:13ag,EldredgeJD:13a,WuZ:16a,Mohseni:17m,EldredgeJD:18a,EldredgeJD:19a,SohnS:20a}]. Many of these studies have been motivated by biological flows, such as flapping/hovering bird flight and fish locomotion, in which the propulsive appendage and/or the animal itself usually performs an oscillatory motion characterized by large incidences. Despite the rather disparate Reynolds number regimes of the physical and modeled problems, e.g. $Re\sim O(10)-O(10^3)$ vs. $Re\rightarrow\infty$, the force predictions from the inviscid flow have achieved impressive accuracy. The implication is that pressure forces due to normal stress dominate in the massively separated flows.

The flow separation at the leading edge (LE) has presented difficulties to models that use point vortices to represent shed vorticity at moderate angles of attack. This has called into question the efficacy of the leading-edge Kutta condition as an appropriate criterion to produce reliable flow simulations. We believe that there are two fundamental issues responsible for those challenges, which we describe as follows. (A) There is always some length and/or time scale at which the topology of isolated point singularities is insufficient to replicate the sectionally holomorphic field established by a continuous vortex sheet. The free-stream flow impinging at the leading edge then has the tendency to penetrate the gaps between the point vortices, which produce a smooth, analytic velocity field everywhere (excluding the points themselves). (B) When the LE spiral is attached to the plate, it effectively translates with that velocity. In other words, the spiral as a whole structure is not exactly `free' in the sense of the Kirchhoff velocity used to convect point vortices. The consequence is that the spiral and the evolution of the circulation distribution within it must be more faithfully represented to produce the `correct' velocity field that keeps the spiral attached to the plate. Due to issues (A) and (B), it can be challenging to even initiate LE shedding without immediately losing the integrity of the sheet to the sweeping component of the free-stream flow. On the other hand, one could argue from a physical basis that, in a viscous fluid, any vorticity generated near the leading edge would remain in an attached boundary layer at moderate incidences~[\onlinecite{EldredgeJD:18a}].

To address the aforementioned issues, point vortex models often employ some \textit{in situ} criteria to control the position, velocity and/or strength of newly shed vortices [\onlinecite{KatzJ:81a,AnsariSA:06b,EldredgeJD:13a,Mohseni:13ag,RameshK:14a,EldredgeJD:19a}]. The most promising of these remedies is the Leading-Edge Suction Parameter (LESP) criterion developed by Ramesh \textit{et al.} [\onlinecite{RameshK:14a}], which suppresses vortex shedding at the LE until a critical LESP value is exceeded and usually corresponds to a critical angle of attack. The physical reasoning behind the LESP is that a finite-thickness airfoil can support some amount of suction as the flow navigates around the (rounded) leading edge without separating. Then, above the critical LESP, the roll-up of the separated flow is reasonably well captured by the release of discrete point vortices that are now `shielded' from a destructive interaction with the free-stream flow. 

However, a third issue of computational cost arises as a practical matter. Namely, the number of points needed to address issue (B) quickly becomes prohibitive. To this end, Darakananda \& Eldredge [\onlinecite{EldredgeJD:19a}] have developed a hybrid model that combines the LESP (issue~A) with a circulation transfer procedure that feeds vorticity from the plate to an attached `sheet' segment of point vortices and finally to an active (i.e. variable circulation) point vortex representing the rolled-up spiral core; each core vortex can be made inactive as it sheds farther downstream. The time-dependent circulation transfer preserves the hydrodynamic impulse and results in an equation of motion for the core vortex (issue B). Moreover, the model can be tuned to achieve a desired balance between the dimensionality and the resolution of flow structures. As a result, dynamically reliable long-time simulations can be obtained with minimal wall-clock time.

The works cited above represent, in part, the state-of-the-art with regard to the robust inviscid modeling of massively separated unsteady aerodynamic flows. In this paper we do not claim nor intend to obtain an improvement over them. Rather, we aim to provide further progress to the mathematical solution for the separated flow at the edges of a flat-plate airfoil in an inviscid fluid that is regularized by a Kutta condition at any incidence. For this purpose, we endeavor to stray as little as possible from the ideal description of the fluid. Namely, we maintain the topological character of continuous vortex sheets and focus our attention on the initial phase of the motion so that self-similar solutions can be leveraged. The fluid is taken as incompressible (with uniform density) as well as irrotational, and since the problem is two dimensional we will utilize the formulation of a complex potential. 

In \S\ref{sec:outer} we discuss the expansion of the outer attached flow around a sharp wedge that will eventually be applied near the leading and trailing edges of the flat plate. The full problem statement of the flow around the plate is given in \S\ref{sec:plate} and which includes a discussion on the constraints of the ensuing inviscid vortex dynamics. This is followed by the derivation of the self-similar approximation and solution in \S\ref{sec:sim}. Section~\ref{sec:results} presents applied results of our model as well as comparisons to Navier-Stokes simulations. The model is then used in \S\ref{sec:lesp} to investigate the separated flow at the leading edge at moderate angle of attack.

\section{The outer flow expansion}\label{sec:outer}
In this section we briefly introduce the concept that will ultimately serve as the basis for our model of the flow around the flat plate. Any analytic function, such as the complex potential, can be expanded in a convergent Laurent series about a point $z_o$ that consists of positive and negative powers of $(z-z_o)$, which correspond to outer and inner expansions, respectively [\onlinecite{Carrier:66a}]. If the function is also analytic at the point $z_o$ itself, then the series becomes the more familiar Taylor series. The outer expansion represents the effect of distant agencies; `Moffatt eddies'~[\onlinecite{MoffattHK:64a}] are a well known flow induced by such phenomena. Conversely, the inner expansion corresponds to local (i.e. singular) agencies; for example, the vorticity field associated with a moving contact line has a dipole distribution [\onlinecite{Mohseni:18p}].

\begin{figure}
\begin{center}
\begin{minipage}{0.32\linewidth}\begin{center}
\includegraphics[width=0.99\textwidth, angle=0]{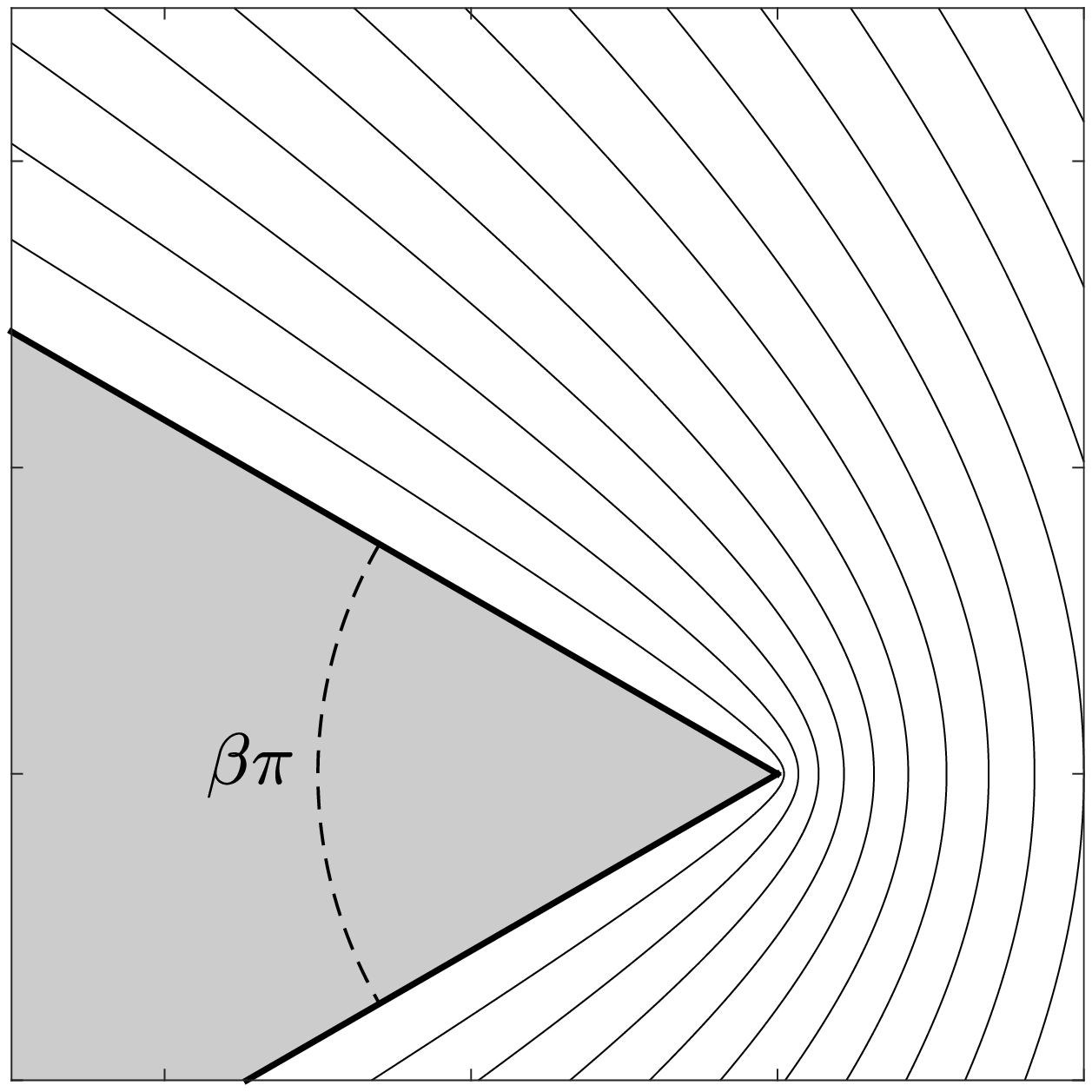} \\(\textit{a})
\end{center}\end{minipage}
\begin{minipage}{0.32\linewidth}\begin{center}
\includegraphics[width=0.99\textwidth, angle=0]{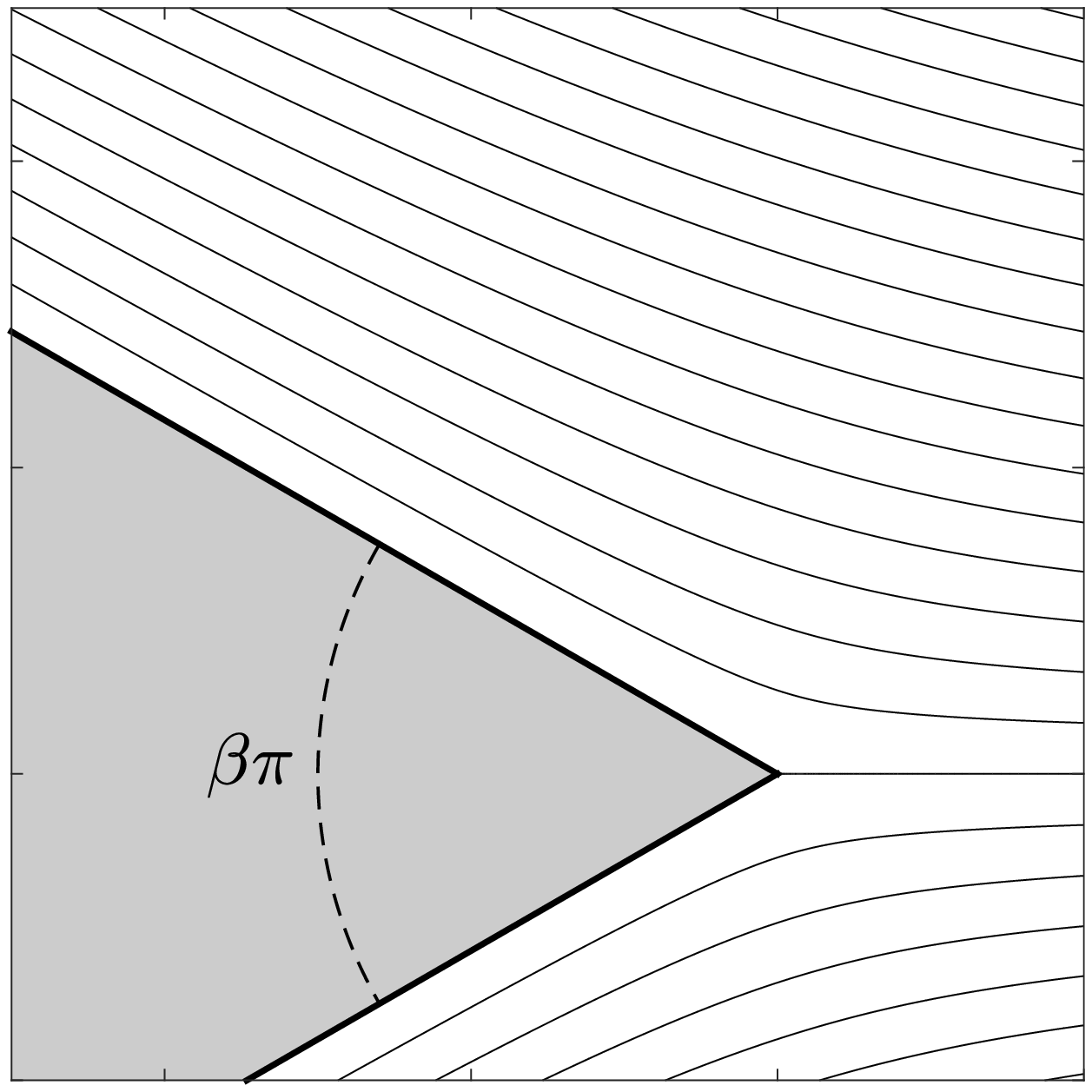} \\(\textit{b})
\end{center}\end{minipage}
\begin{minipage}{0.32\linewidth}\begin{center}
\includegraphics[width=0.99\textwidth, angle=0]{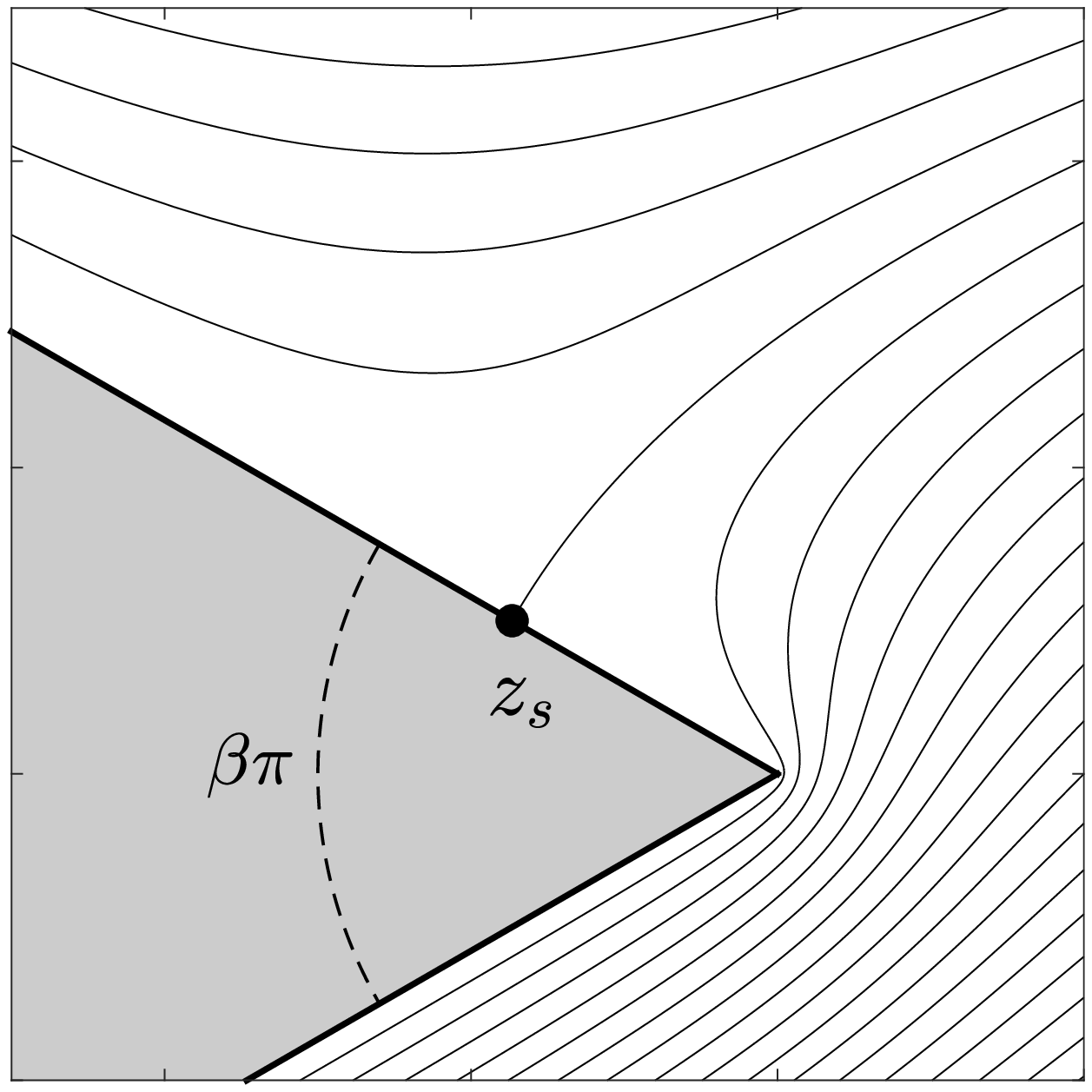} \\(\textit{c})
\end{center}\end{minipage}
\caption{Example streamlines of attached flows around a wedge of angle $\beta\pi$: (\textit{a}) singular flow from bottom-to-top, (\textit{b}) higher-order regular flow from left-to-right, and (\textit{c}) the composite flow of equation~(\ref{eqn:outer}). In (\textit{c}) the black dot marks the stagnation point $z_s$ on the wedge surface.}
\label{fig:stagpt}
\end{center}
\end{figure}

In the particular case of flow around a sharp wedge of interior angle $\beta\pi$ ($0\leq\beta<1$) with apex located at $z=0$ (see figure~\ref{fig:stagpt} for geometry), the complex potential of the attached flow can be written as [\onlinecite{Mohseni:20a}]
\begin{equation}\label{eqn:Laurent}
W_a(z,t)=A_0(t)+A_1(t)z^n+A_2(t)z^{2n}+\dots+A_k(t)z^{kn}+\dots ,
\end{equation}
where $n=1/(2-\beta)$ and while $W_a$ is finite at $z=0$, it is not analytic since its complex derivative does not exist there. Also note that (\ref{eqn:Laurent}) has set to zero the higher-order singularities of the inner expansion corresponding to $k<0$ terms. This expansion is assumed to be valid near the apex at small times after the onset of motion. As such, the series is usually truncated after $k=1$ since this yields the dominating, singular term in the velocity expression $\md{W_a}/\md{z}$ of this attached flow. Hence, the physical geometry reintroduces an inner expansion term of fractional power $-(1-n)$ into the velocity. Pullin~[\onlinecite{Pullin:78a}] regularized this flow by allowing separation at the apex to shed a self-similar conventional vortex sheet. Those solutions are sometimes used as an initial condition to begin inviscid simulations with arbitrary unsteadiness (e.g. Refs. \onlinecite{JonesMA:03a},\onlinecite{SohnS:20a}). In DeVoria \& Mohseni~[\onlinecite{Mohseni:20a}] we significantly augmented the solution space by considering an entrainment boundary condition (in place of no through-flow) that allows a complex coefficient $A_1\in\mathbb{C}$ and the shedding of a vortex-entrainment sheet [\onlinecite{Mohseni:19d}]. 

In the present paper, we adopt the standard zero entrainment boundary condition of no through-flow. However, the novelty of the study is in consideration of the higher-order terms in the expansion (\ref{eqn:Laurent}). Since $\frac{1}{2}\leq n< 1$, the $k\geq 2$ terms are regular at the apex, each having a defined derivative there, and again represent the outer expansion due to distant agencies. For simplicity, we truncate the series after $k=2$ and to satisfy zero entrainment on the wedge faces, where $arg\{z^n\}=n\theta=\pm\pi/2$, we write $A_1=-ia$ and $A_2=b\in\mathbb{R}$. Also, $A_0$ can be set to zero without affecting the attached flow velocity field, which becomes
\begin{equation}\label{eqn:outer}
\der{W_a}{z}=nz^{n-1}\big\{-ia+2bz^n\big\}.
\end{equation}
As mentioned in Ref.~[\onlinecite{Mohseni:20a}], the higher-order terms in (\ref{eqn:Laurent}) introduce additional topological features. In the case of (\ref{eqn:outer}), the regular flow opposes the singular flow on one wedge face and contributes to it on the other face, and so a hyperbolic stagnation point $z_s$ appears on the surface of opposing flows at a radial distance from the apex of
\begin{equation}
r_s=\left|\frac{a}{2b}\right|^{1/n},
\end{equation}
where $z_s$ is on the upper surface if $a/(2b)>0$ and on the lower surface if $a/(2b)<0$. Figures~\ref{fig:stagpt}(\textit{a})--(\textit{c}) show examples of the attached flow streamlines for the singular, regular, and composite flows. The radius $r_s$ is a length scale intrinsic to the composite flow and evidently indicates the `proximity' of the distant agency of strength $b$ to the apex where the potential of strength $a$ dominates. In \S\ref{sec:sim} we will show that equation (\ref{eqn:outer}) yields the asymptotic outer flow near the leading and trailing edges of a flat plate ($n=1/2$) of chord length $c$ that is translating with velocity $U(t)$ at an angle of attack $\alpha$. More specifically, we will find that $a=c^{1/2}U\sin\alpha$ and $b=U\cos\alpha$, and that $r_s$ indicates the interaction between the leading-edge and trailing-edge flow developments. Note that $b=U\cos\alpha$ is the sweeping component of the free-stream flow. 

\begin{figure}
\begin{center}
\begin{minipage}{0.7\linewidth}\begin{center}
\includegraphics[width=0.99\textwidth, angle=0]{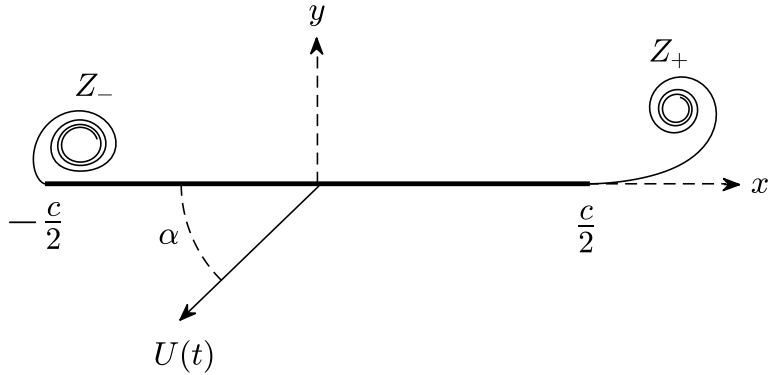}%
\end{center}\end{minipage}
\caption{Definition of the $x$-$y$ coordinate system instantaneously coinciding with the mid-chord of the flat plate that translates with speed $U(t)$ and angle of attack $\alpha$. The trailing edge is at $x=c/2$, while the leading edge is at $x=-c/2$. The complex positions of the corresponding shed vortex sheets are $Z_{+}$ and $Z_{-}$.}
\label{fig:plate}
\end{center}
\end{figure}
\section{The flat-plate airfoil}\label{sec:plate}
Here, we give the problem formulation for the general flow around a plate that translates with speed $U(t)$ at constant angle of attack $\alpha$. In the inertial frame with the fluid at rest at infinity, the plate motion is to the left and downward with velocity $U_b+iV_b=-U(t)\e{i\alpha}$ as shown in figure~\ref{fig:plate}. As usual, the flat-plate geometry with chord length $c$ in the physical plane, where $z=x+iy$ is the complex coordinate, is obtained from the mapping of a circular boundary with radius $c/4$ in a virtual $\zeta$-plane via the Joukowski transformation:
\begin{equation}\label{eqn:Jtrans}
\zeta(z)=\tfrac{1}{2}\big(z+\sqrt{z^2-c^2/4}\big).
\end{equation}
The origin of the coordinate system instantaneously coincides with the mid-chord of the plate so that its trailing and leading edges are located at $z=+c/2$ and $z=-c/2$, respectively. The vortex sheets shed from each edge have corresponding positions $Z_{\pm}$ in the physical plane and $\zeta_{\pm}=\zeta(Z_\pm)$ in the virtual plane.

In the next section we give the expression for the total complex potential that describes the flow. The scalar potential is $\phi$ and the stream function is $\psi$, both of which satisfy the Laplace equation in the fluid domain $\Omega_f$ with appropriate boundary conditions. Again, the topological character of continuous vortex sheets will be maintained so as to circumvent any issues specific to point vortex methods.

\subsection{The complex potential}\label{sec:W}
The total complex potential only exists in the analytic region $\Omega_f$, i.e. the fluid domain, that is outside the plate and any sheets of discontinuity shed from the plate. Let the contour immediately surrounding these inner boundaries be $C_i\in\Omega_f$. Since the spatial domain is two dimensional, then a discontinuity in $W_o$ may also exist due to a logarithmic constituent and requires a branch cut $C_{cut}$ to be specified so that the logarithm is uniquely defined. The cut intersects the plate at a point $z_c(t)$ and must extend to infinity where it connects to an all-enclosing contour $C_\infty$. Each side of $C_{cut}$ is a distinct portion of the fluid boundary. This ensures that $W_o$ is single-valued in the simply connected region defined by the \textit{complete} fluid boundary $\partial\Omega_f=C_i+C_{cut}+C_\infty$. Therefore, all the usual statements regarding the properties of analytic functions apply in the region $\Omega_f$ (e.g. see Refs. [\onlinecite{TownsendEJ:42a,ChurchillRV:48a,Carrier:66a}]). 

The real and imaginary parts of the jump in $W_o$ (due to the logarithm) across the cut respectively correspond to the net circulation $\Gamma_o(t)$ around and the net flux $Q_o(t)$ across a contour $C_o$ that encloses the plate and any shed sheets~[\onlinecite{Mohseni:19d}]. As a consequence of the Cauchy-Goursat theorem, we must have $\Gamma_o=\Gamma_i=\Gamma_\infty$, where $\Gamma_i$ and $\Gamma_\infty$ are the circulations around the (closed) contours $C_i$ and $C_\infty$, respectively. In other words, the arbitrary closed contour $C_o$ is reconcilable within the (topologically) annular region between $C_i$ and $C_\infty$, and thus the circulation around any such contour is $\Gamma_o$. Analogous statements can be made about the flux $Q_o$.

The quantities $\Gamma_o$ and $Q_o$ must be specified as input, or otherwise determined, to have a unique solution for the complex potential. Since we have imposed the normal boundary condition of zero entrainment everywhere on the inner boundary $C_i$, then $Q_o\equiv 0$ necessarily. In the case of a flow started from rest (when the fluid is free of discontinuities), Kelvin's circulation theorem is usually invoked to arrive at the result that $\Gamma_o(t)\equiv 0$. It is often taken for granted that this result requires the assumptions that $C_o$ does not, at any time, intersect a discontinuity in velocity nor pressure, and that no non-conservative forces $\vect{f}$ (with $\curl{f}\neq\mathbf{0}$) act tangentially anywhere along the contour. Such non-conservative forces acting on the portion of $C_i$ that coincides with the plate surface could be used to model vorticity generation. While we will not explore the physical details of any of these circulation-generating mechanisms in the current study, we will carry through the analysis with $\Gamma_o(t)\neq 0$ for general posterity.

The total circulations in the shed sheets $Z_{+}$ and $Z_{-}$ are $\Gamma_{+}$ and $\Gamma_{-}$, respectively, while the `bound' circulation around the plate is $\Gamma_b$. Accordingly, the circulation around the arbitrary contour $C_o$ enclosing the entire plate/sheet system is $\Gamma_o=\Gamma_b+\Gamma_{+}+\Gamma_{-}$. In general, so long as the boundary $\partial\Omega_f$ is known along with the value of $W_o$ on $\partial\Omega_f$, then a unique solution exists for $W_o$ with a given value of $\Gamma_o$ and without requiring any further information about the bound or vortex sheet circulations. However, we have supposed that the vortex sheets emanate from the edges of the plate on the physical basis that the velocity remains finite there. Hence, two more relations are required to ensure that this is indeed satisfied. These supplemental constraints are the Kutta conditions and are sufficient to determine the individual circulations $\Gamma_{+}$ and $\Gamma_{-}$, and since $\Gamma_o$ will be known from the dynamics (via Kelvin's theorem), then $\Gamma_b$ is obtained as well. This completes the unique determination of the complex potential and we may now state its analytic expression.

The total complex potential can be written as $W_o=W_b+W_{+}+W_{-}$, where $W_b$ and $W_\pm$ are the contributions from the plate and the shed vortex sheets. Let $W_\infty=-zU\e{-i\alpha}$ be the potential of a moving reference frame translating with the plate. By the circle theorem~[\onlinecite{MilneThomsonLM:68a}], the plate potential $W_b$ can be represented in closed form in this non-inertial frame by the image systems of $W_\pm$ and $W_\infty$. The resulting expression for this relative complex potential defined as $W\equiv W_o-W_\infty$ is:
\begin{subequations}
\begin{eqnarray}
W &=& W_a+W_v+\frac{\Gamma_o}{2\pi i}\log\zeta  \label{eqn:W} \\
W_a &=& zU\cos\alpha-iU\sin\alpha\sqrt{z^2-c^2/4} \label{eqn:Wa} \\
W_v &=&\frac{1}{2\pi i}\sum_{\pm}\int_0^{\Gamma_{\pm}}\bb{\log\lrp{\zeta-\zeta_{\pm}}-\log\lrp{\zeta-\zeta^{i}_{\pm}}}\mdi{\Gamma}, \label{eqn:Wv} 
\end{eqnarray}
\end{subequations}
where $W_a$ is the potential of the attached flow, $W_v$ is the potential due to the vortex system and includes the contributions from the sheets in the fluid as well as their corresponding images located at $\zeta^{i}_\pm=(c/4)^2/\cc{\zeta}_\pm$; an overbar denotes the complex conjugate. Here and elsewhere, the sum is shorthand for the addition of the integrals with terms respectively having ($+$) and ($-$) subscripts. The logarithmic constituent is the image of the net circulation $\Gamma_o$ at infinity, which here can be written explicitly as the uniformly valid term $\log\zeta$ via the circle theorem. This circulation does not affect the kinematics represented by the normal boundary condition on the plate (Neumann problem). In fact, it is the single remaining degree of freedom in the \textit{tangential} boundary condition (Dirichlet problem), which represents the dynamics. As mentioned above, the physical principle used to fix the value of $\Gamma_o$ is Kelvin's theorem. 
Having specified the flow solution, next we discuss the expression for the force exerted on the plate.

\subsection{The force on the plate}\label{sec:force}
The force on the plate is computed as the integral of pressure around its surface. Since the vortex sheets are assumed to have a zero pressure jump, then the integral can be augmented to include those sheets and thus the contour replaced by that of the inner boundary $C_i$. Following Newman~[\onlinecite{NewmanJN:77a}] and temporarily using vector notation, this pressure integral can be rewritten to give the force as
\begin{equation}\label{eqn:F1}
\vect{F}=\rho\der{}{t}\oint_{C_i}\phi\uv{n}\mdi{l}+\rho\oint_{C_\infty}\lrb{\frac{1}{2}|\vect{u}|^2\uv{n}-\vect{u}(\vect{u}\cdot\uv{n})}\md{l}.
\end{equation}
From here, there are two different ways to proceed, the details of which depend on the reference frame; the end result is the same, of course. In the \textit{moving} frame, the $C_\infty$ integral yields the familiar Joukowski force associated with $\Gamma_o$ (see Batchelor~[\onlinecite{Batchelor:67a}], pp. 406-7). In the \textit{stationary} frame the integral vanishes, but the Joukowski force reappears from the time derivative of the logarithmic constituent of $\phi$ in the $C_i$ integral, which we discuss next.

In either frame, due to the zero entrainment boundary condition, the scalar potential in~(\ref{eqn:F1}) can be replaced by the complex potential. Generally, an integral of the stream function $\psi$ is also present, but which is proportional to the fluid mass displaced by the body~[\onlinecite{YihCS:69a}], and so is zero for the flat plate. Opting for the moving frame, the residue theorem can be used to replace $C_i$ with $C_\infty$ for $W_a$ and $W_v$ in (\ref{eqn:W}). For the log term of $W$, however, we must use the more general Cauchy-Goursat theorem (from which the residue theorem derives) that involves the complete fluid boundary. Thus the replacement contour includes the cut, $C_\infty+C_{cut}$, and it is easy to show that the integral of $(\Gamma_o/2\pi i)\log\zeta$ over this contour evaluates to $\Gamma_o z_c$, where again $z_c$ is the position where the cut intersects the plate. In the stationary frame, with coordinates $\tilde{z}$  and $\tilde{\zeta}$ say, the same result is obtained since the logarithm becomes $\log(\tilde{\zeta}-\tilde{\zeta_o})$, where $\tilde{\zeta_o}(t)$ is the position of the plate mid-chord. However, the time derivative of $z_c$ in this frame is $\md{z_c}/\md{t}+(-U\e{i\alpha})$ and the Joukowski force is recovered from the latter term. The first term is the velocity of $z_c(t)$ on the plate surface in the moving frame.

Next, we decompose the total force as 
\begin{subequations}
\begin{equation}
F_t+iF_n = \underbrace{F_t^{(v)}+iF_n^{(v)}}_\text{vortex force} + \underbrace{F_t^{(b)}+iF_n^{(b)}}_\text{body force},
\end{equation}
where for each force constituent $F_t^{(\cdot)}$, $F_n^{(\cdot)}$ are the components tangential and normal to the plate, respectively. The `vortex force' is due to $W_v$ and the `body force' combines the contributions from $W_a$ and $\Gamma_o$, and these are expressed by the following:
\begin{eqnarray}
F_t^{(v)}+iF_n^{(v)} &=& i\rho\der{}{t}\left\{\sum_\pm\int_0^{\Gamma_\pm}\lrp{\zeta_\pm-\zeta_\pm^i}\mdi{\Gamma}\right\} \label{eqn:Fv} \\
F_t^{(b)}+iF_n^{(b)} &=& i\rho c^2\frac{\pi}{4}\der{U}{t}\sin\alpha + i\rho\der{}{t}\bp{z_c\Gamma_o}+i\rho\Gamma_o\bp{-U\e{i\alpha}}. \label{eqn:Fa}
\end{eqnarray}
\end{subequations}
The conventional definitions of the lift and drag forces are $D+iL=(F_t+iF_n)\e{-i\alpha}$. Next, we briefly examine a revealing connection between the plate force and the dynamics of the shed vortex system.

\subsection{Inviscid vortex dynamics and impulse invariant}\label{sec:impulse}
For an inviscid fluid of infinite extent there are flow invariants associated with integrals of the vorticity field [\onlinecite{Batchelor:67a,Saffman:92a}] that represent the net amounts of vorticity (or circulation in 2D), impulse and kinetic energy. Here, we are interested in the delivery of impulse to the fluid. For our problem the fluid remains irrotational and the effect of vorticity is represented by vortex sheets, which are boundaries to the fluid domain. As such, the impulse delivered to the fluid is equal to that of the freely-shed vortex sheets. Moreover, since these sheets do not support a pressure jump, then the \textit{net fluid impulse} can only be increased (or decreased) by an increment (or decrement) delivered by the plate, i.e. only if it accelerates. 

In the case that $\Gamma_o\equiv 0$, then from (\ref{eqn:Fa}) we see that the plate can only deliver a net impulse in the direction normal to itself and therefore the vortex force in (\ref{eqn:Fv}) (due to the impulse of the shed vortex system) will only generate a non-zero plate-normal component. This means that the plate-tangential forces of the leading- and trailing-edge vortex systems must mutually cancel by the movement and circulation dynamics of each sheet. This is despite any visual asymmetry of the total vortex system. These claims will be validated in \S\ref{sec:results} with numerically computed solutions.


\section{Approximate solutions near the plate edges}\label{sec:sim}
Pullin \& Wang [\onlinecite{Pullin:04b}] approached this problem using a perturbation expansion of the singular-order similarity solutions for the separated flow around a semi-infinite flat plate (i.e. with no intrinsic length scale) as computed by Pullin [\onlinecite{Pullin:78a}]. Expansions of both the vortex sheet positions and circulations were carried out in the small parameter $\epsilon(t)=\sqrt{R_v(t)/c}$, where $R_v(t)$ is the time-dependent length scale appropriate to the size of the growing vortex spiral and again $c$ is the chord length of the plate. Recently, Sohn [\onlinecite{SohnS:20a}] used the same approach to include rotation of the plate. Interestingly, rotation does not change the fundamental form of the asymptotic equation as the effect is embedded within $R_v(t)$, which is known \textit{a priori} to the solution. Hence, if desired we can include the effect of rotation in the similarity problem by using Sohn's modified definition of $R_v$.

The objective of Pullin \& Wang [\onlinecite{Pullin:04b}] was to capture the effect of asymmetry due to the sweeping motion of the free-stream component parallel to the plate, that is $U\cos\alpha$. Again, this was implemented through higher-order corrections to the sheet positions and circulations, with the expectation that subsequent corrections to the forces on the plate would be obtained. This led to a series of ordered integro-differential equations to be solved in succession, the first being the singular-order equation of Pullin [\onlinecite{Pullin:78a}]. The approach was successful for the former quantities. However, due to the construction of the expansions, the higher-order corrections to the vortex force vanished through cancellations. 

Here, we consider a similar approach, but which does not employ an infinite-term perturbation series for the sheet positions and circulations. Instead, under certain conditions to be identified below (see (\ref{eqn:conds})), the asymmetric sweeping effect can be brought into the same order as the singular attached flow around the edges. As mentioned in \S\ref{sec:outer}, the result corresponds to the expansion of the outer flow to higher order. In essence, a new singular-order governing equation is obtained that includes the asymmetry, but which is only trivially altered from the integro-differential equation solved in Pullin [\onlinecite{Pullin:78a}].

First, we write the vortex sheet locations $Z_\pm$ and the total circulations $\Gamma_\pm$ as
\begin{subequations}
\begin{eqnarray}
Z_{+}&=&\frac{c}{2}\big(1+2\epsilon^2\omega_{+}\big),\quad\quad \Gamma_{+}=J_{+}G  \label{eqn:wp}\\   
Z_{-}&=&-\frac{c}{2}\big(1+2\epsilon^2\cc{\omega}_{-}\big),\quad\quad \Gamma_{-}=-J_{-}G \label{eqn:wn},  
\end{eqnarray}
\end{subequations}
where $\omega_{\pm}$ are the corresponding non-dimensional self-similar shape functions, and again $\epsilon^2 =R_v/c$. Note the complex conjugate of $\omega_{-}$ defining $Z_{-}$ in (\ref{eqn:wn}). Also, $J_\pm$ are the non-dimensional circulation \textit{magnitudes} with $G(t)$ as the temporal growth of the circulation scaling. Further requirements of similarity restrict the plate velocity to be of the form
\begin{equation}
U(t)=Bt^m,
\end{equation} 
where $m$ is given; we shall call this quantity the acceleration exponent. We refer the reader to Refs. [\onlinecite{Pullin:78a}] and [\onlinecite{Mohseni:20a}] for more details on the self-similar problem setup. It can be shown that the velocity induced by one vortex sheet on the other at the opposing edge is of the same order as terms that we eventually will ignore and is therefore neglected now. Next, we give some detail on the case of the trailing-edge sheet $Z_{+}$, for the results applicable to the leading-edge sheet $Z_{-}$ follow immediately with minor changes.

The velocity field is obtained by differentiating (\ref{eqn:W}) with respect to $z$ and using $\md{W}/\md{z}=(\md{\zeta}/\md{z})\md{W}/\md{\zeta}$ where appropriate. When $z\rightarrow Z_{+}$ and $\zeta\rightarrow \zeta_{+}$ the result becomes the Birkhoff-Rott equation, $\partial \cc{Z}_{+}/\partial t=\md{W}/\md{z}$, which is the (kinematic) governing equation for the sheet. A careful expansion in $\epsilon$ of the right-hand side yields
\begin{subequations}
\begin{eqnarray}\label{eqn:dWdz}
\der{W}{z}&=&\frac{\dot{R}_v}{C_o}\left[\frac{1}{2\sqrt{\omega_{+}}}\left\{-i\left(1+\frac{\Gamma_o}{\pi cU\sin\alpha}\right)+(2\epsilon\cot\alpha)\sqrt{\omega_{+}}+I_0\right\}+\epsilon\big\{I_0-I_1\big\}\right],
\end{eqnarray}
where $C_o=\epsilon\dot{R}_v/(U\sin\alpha)$ and
\begin{eqnarray}
I_0&=&\frac{J_{+}}{2\pi i}\int_0^1\left[\frac{1}{\sqrt{\omega_{+}}-\sqrt{\omega_{+}^\prime}}-\frac{1}{\sqrt{\omega_{+}}+\sqrt{\cc{\omega}_{+}^\prime}}\right]\md{\lambda^\prime}\\
I_1&=&\frac{J_{+}}{2\pi i}\int_0^1\frac{1}{2\sqrt{\omega_{+}}}\left[\frac{\omega_{+}-\omega_{+}^\prime}{\left(\sqrt{\omega_{+}}-\sqrt{\omega_{+}^\prime}\right)^2}-\frac{\omega_{+}-\cc{\omega}_{+}^\prime}{\left(\sqrt{\omega_{+}}+\sqrt{\cc{\omega}_{+}^\prime}\right)^2}\right]\md{\lambda^\prime},
\end{eqnarray}
\end{subequations}
where $\lambda=1-\Gamma/(JG)$ is the dimensionless circulation similarity variable along the sheet and $G(t)=\epsilon cU\sin\alpha$; a prime indicates a dummy variable for integration. Next, substitution of (\ref{eqn:wp}) into the left-hand side of the governing equation gives
\begin{equation}\label{eqn:dzdt}
\pder{\cc{Z}_{+}}{t}=\dot{R}_v\left[\cc{\omega}_{+}+Q(1-\lambda)\der{\cc{\omega}_{+}}{\lambda}\right],
\end{equation}
with $Q=(4m+1)/(2m+2)$. This equation has no explicit $O(\epsilon)$ term since we did not consider a series expansion of $Z_{+}$. At first glance, it seems we must then neglect all of the $O(\epsilon)$ terms in (\ref{eqn:dWdz}). However, we note that the angle of attack, which is an independently specified parameter, appears in combination with $\epsilon$ in the $2\cot\alpha$ term. Therefore, we impose the constraints
\begin{equation}\label{eqn:conds}
\epsilon\ll 1,\quad\quad\quad 2\epsilon\cot\alpha\equiv \eta_\alpha\sim O(1),
\end{equation}
and drop the terms in the second set of curly brackets in (\ref{eqn:dWdz}). We are technically restricted to a moderate angle of attack range, as well as small times through $\epsilon(t) =\sqrt{R_v(t)/c}$, which represents the finite length scale $c$. The ranges of validity will be investigated \textit{a~posterori} with numerically computed solutions. The $2\cot\alpha$ term represents the asymmetric sweeping effect and can be found in the $O(\epsilon)$ equation of Pullin \& Wang [\onlinecite{Pullin:04b}]. From (\ref{eqn:dWdz}) and (\ref{eqn:outer}) we readily see that the attached flow terms (with $\Gamma_o=0$) yield $a=c^{1/2}U\sin\alpha$ and $b=U\cos\alpha$ as stated in \S\ref{sec:outer}. As such, $\eta_\alpha$ reflects the interaction between the flow developments at the leading and trailing edges of the plate. 

Equation (\ref{eqn:dzdt}) remains unchanged from Refs. [\onlinecite{Pullin:78a,Pullin:04b,Mohseni:20a}], and following those works we take $C_o=1/2$. Next we write $\Gamma_o=J_o(\pi cU\sin\alpha)$ so that $J_o$ is the fraction of the steady-state circulation of classical aerodynamics. We note that since $\epsilon(t)$ is a function of time, then so too is $\eta_\alpha(t)$, and likewise for $J_o(t)$ due to $\Gamma_o(t)$ and $U(t)$. As such there is a separate, implicit time scale in the physical domain solution as $Z_{+}(\Gamma,t)=c/2+c\epsilon^2(t)\omega_{+}(\lambda,\eta_\alpha(t),J_o(t))$ and $\Gamma_{+}(t)=G(t)J_{+}(\eta_\alpha(t),J_o(t))$. We will utilize this to construct a more complex evolution of the flow structures; the procedure for this task will be further explained shortly.

The non-dimensional governing equation for $\omega_{+}$ is given by equating (\ref{eqn:dWdz}) and (\ref{eqn:dzdt}), then applying (\ref{eqn:conds}). The equation for the leading-edge sheet $\omega_{-}$ is obtained from the same equation, but with $\eta_\alpha\rightarrow-\eta_\alpha$ and $J_o\rightarrow -J_o$. In this case, the solution can be transformed to the physical leading-edge location by (\ref{eqn:wn}). Regarding the physical significance of $\eta_\alpha$, we need only consider positive values, which correspond to a positive angle of attack relative to the left-to-right uniform flow in the axes fixed on the plate. On the other hand, $J_o$ may take either sign in accordance with the sign of $\Gamma_o$. As such, we finally obtain the governing equations for $\omega_{\pm}$ as:
\begin{subequations}\label{eqn:GE}
\begin{gather}
\cc{\omega}_{\pm}+Q(1-\lambda)\der{\cc{\omega}_{\pm}}{\lambda}=\der{\Omega}{\omega} \\
\der{\Omega}{\omega}=\frac{1}{\sqrt{\omega_{\pm}}}\left\{-i(1\pm J_o)\pm \eta_\alpha\sqrt{\omega_{\pm}}+\frac{J_{\pm}}{2\pi i}\int_0^1\left[\frac{1}{\sqrt{\omega_{\pm}}-\sqrt{\omega_{\pm}^\prime}}-\frac{1}{\sqrt{\omega_{\pm}}+\sqrt{\cc{\omega}_{\pm}^\prime}}\right]\md{\lambda^\prime}\right\}, 
\end{gather}
\end{subequations}
where $\Omega(\omega)$ is the non-dimensional complex potential. Note that the ($\pm$) signs are taken individually for $\omega_{+}$ and $\omega_{-}$, respectively, and no summation is implied. The corresponding Kutta conditions at each edge are:
\begin{equation}\label{eqn:Kutta}
(1\pm J_o)-\frac{J_{\pm}}{2\pi}\int_0^1\left[\frac{1}{\sqrt{\omega_{\pm}}}+\frac{1}{\sqrt{\cc{\omega}_{\pm}}}\right]\mdi{\lambda}=0.
\end{equation}
Upon specification of $J_o$ and $\eta_\alpha$, the equations can be solved for $\omega_{\pm}$ and $J_{\pm}$. In particular, when $\eta_\alpha=J_o=0$, (\ref{eqn:GE}) reduces to the equation numerically solved by Pullin [\onlinecite{Pullin:78a}] and thus inclusion of non-zero values of these parameters requires only minor amendment. We use the numerical scheme described in DeVoria \& Mohseni [\onlinecite{Mohseni:20a}], which studied cases of non-zero entrainment in the starting-flow separation over sharp wedges.

\begin{figure}
\begin{center}
\begin{minipage}{0.99\linewidth}\begin{center}
\includegraphics[width=0.99\textwidth, angle=0]{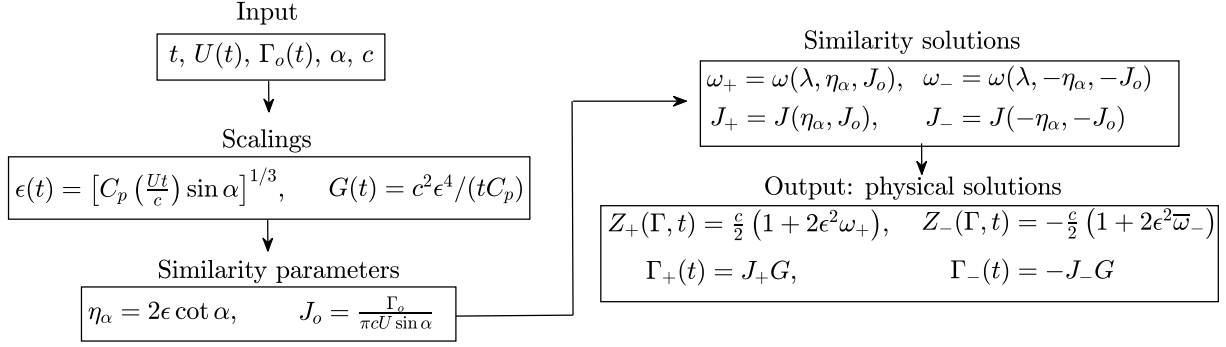}%
\end{center}\end{minipage}
\caption{Flow chart detailing the transformation of the self-similar solutions to the physical space for given input data. The similarity parameters are given to the governing equation in (\ref{eqn:GE}) with an appropriate initial condition.}
\label{fig:chart}
\end{center}
\end{figure}
Once the self-similar solutions have been computed for a given pair of $\eta_\alpha$ and $J_o$, they can be transformed to physical solutions using (\ref{eqn:wp}) and (\ref{eqn:wn}) with the following scalings:
\begin{equation}\label{eqn:scale}
\epsilon(t)=\left[C_p\left(\frac{Ut}{c}\right)\sin\alpha\right]^{1/3},\quad\quad\quad G(t)=\frac{c^2\epsilon^4}{tC_p}, 
\end{equation}
where $C_p=3/[4(1+m)]$ is a constant for given $m$. The transformation process is detailed by the flow chart in figure~\ref{fig:chart}. As mentioned earlier, the result will include the implicit time scale contained in the parameters $\eta_\alpha$ and $J_o$, which represent two separate asymmetric effects. A given trajectory or curve in the $\eta_\alpha$-$J_o$ similarity space (parameterized by time) is obtained by beginning from the baseline and incrementing the values along the curve using the previous solution as an initial condition to the governing equation (\ref{eqn:GE}). We used an analogous procedure in Ref. [\onlinecite{Mohseni:20a}] to represent a time-dependent shedding angle of a vortex-entrainment sheet separating from a non-cusped wedge.

\subsection{Similarity results}
In this section we present results from two sets of self-similar solutions corresponding to simple combinations of the parameters $\eta_\alpha$ and $J_o$. We will see that the significant effect of asymmetry is indeed captured. Physical airfoil solutions will be discussed in \S\ref{sec:results}.

\begin{figure}
\begin{center}
\begin{minipage}{0.49\linewidth}\begin{center}
\includegraphics[width=0.99\textwidth, angle=0]{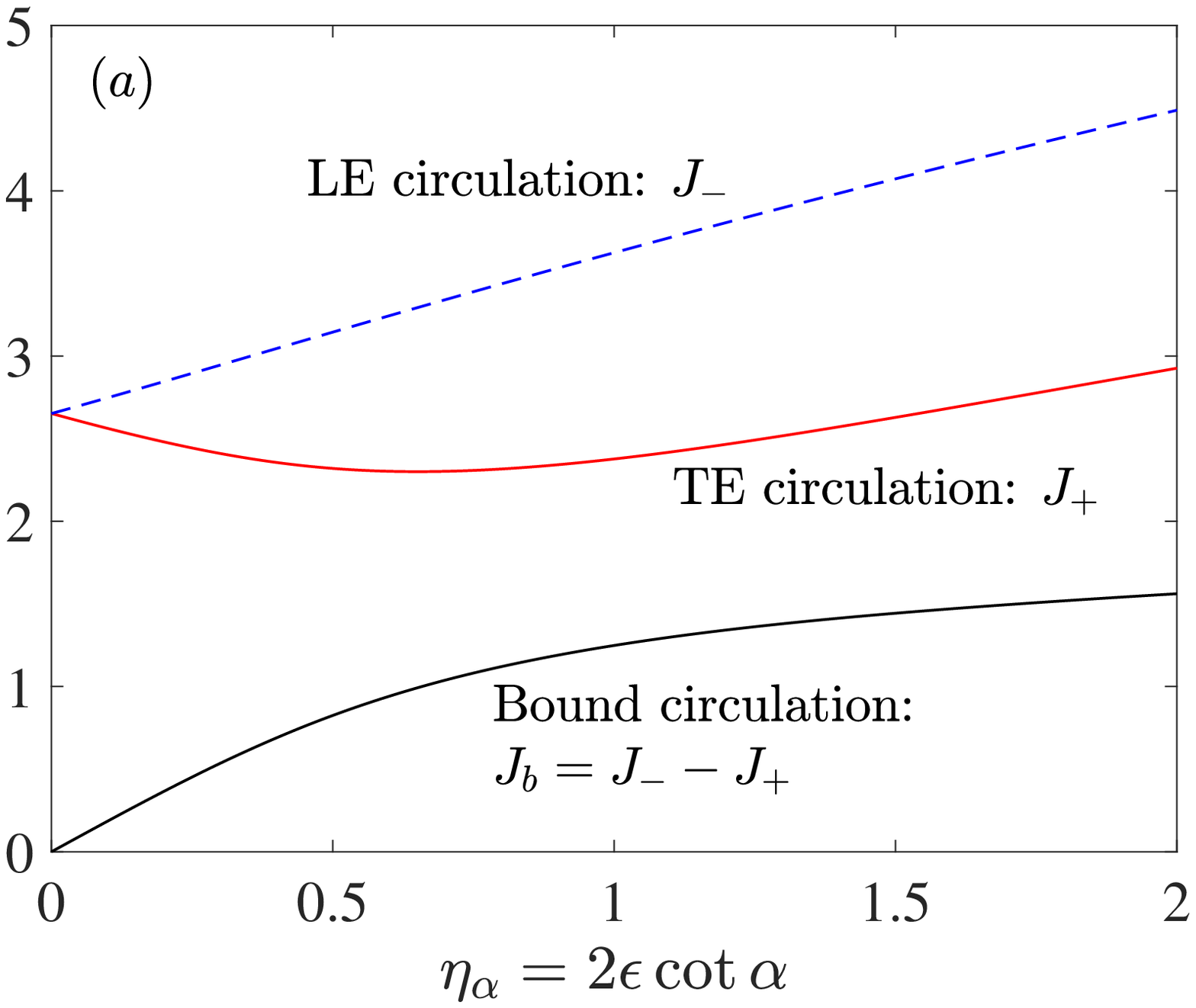}
\end{center}\end{minipage}
\begin{minipage}{0.49\linewidth}\begin{center}
\includegraphics[width=0.99\textwidth, angle=0]{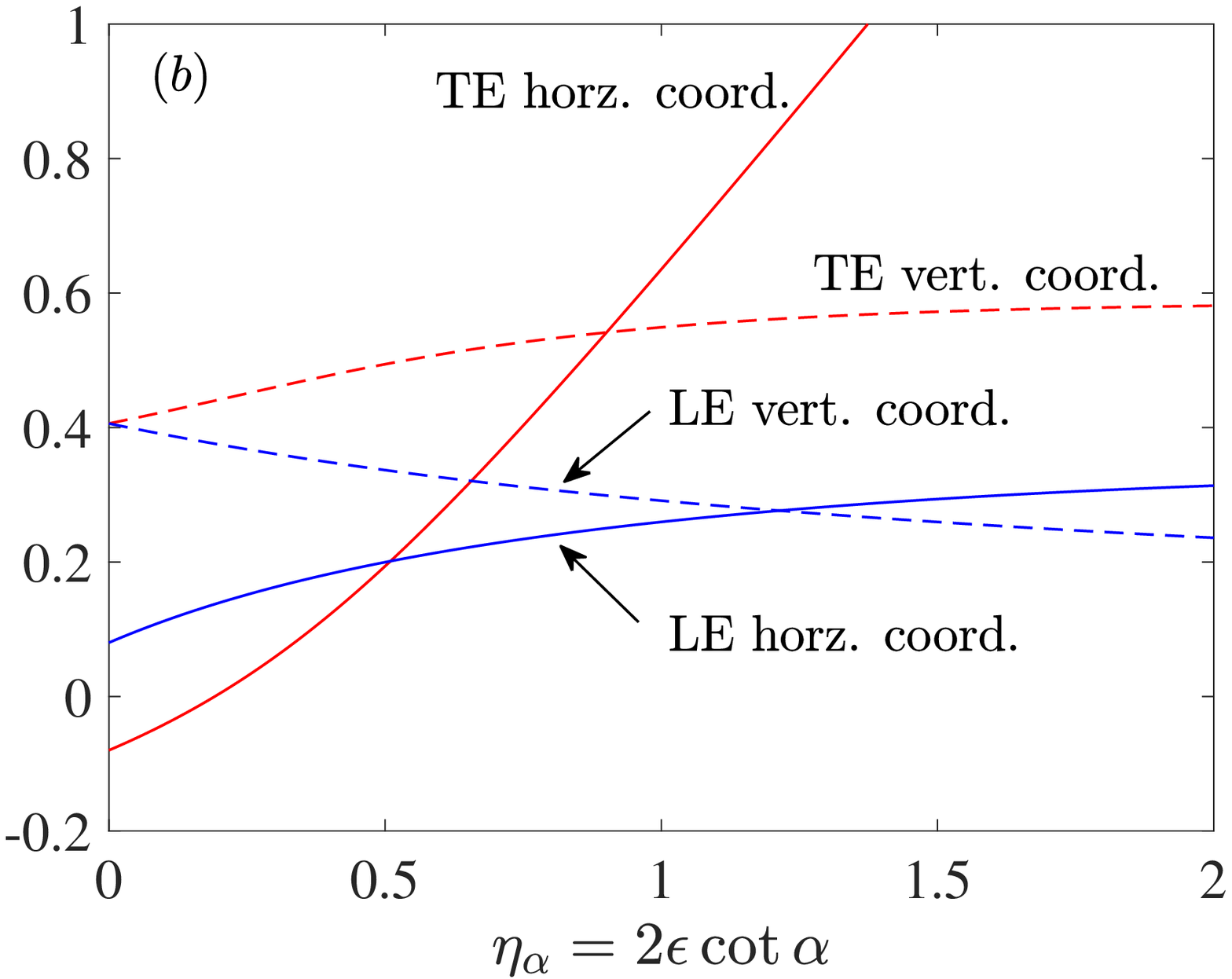}
\end{center}\end{minipage}
\caption{Variation of similarity results with $\eta_\alpha$ for trailing edge $(+)$ and leading edge $(-)$ solutions. The parameters are $m=0$ (impulsive acceleration) and $J_o=0$. (\textit{a}) Non-dimensional circulation magnitudes in the sheets and the bound circulation. (\textit{b}) Horizontal and vertical coordinates of the vortex spiral core locations for each sheet; also see figure~\ref{fig:mk}(\textit{a}).}
\label{fig:Jcore}
\end{center}
\end{figure}

The first set of solutions is for varying $\eta_\alpha$ with $J_o=0$. Figure~\ref{fig:Jcore}(\textit{a}) plots the non-dimensional circulation magnitudes of the leading-edge (LE) and trailing-edge (TE) vortex sheets, $J_{-}$ and $J_{+}$, respectively, for the case of an impulsively accelerated plate $m=0$. The constant acceleration case, $m=1$, displays the same trends, albeit with different quantitative values. For the LE sheet, the circulation increases with $\eta_\alpha$, while it initially decreases for the TE sheet. This is the same trend reported by Pullin \& Wang [\onlinecite{Pullin:04b}]. In fact, by making the linear approximations $J_{\pm}=J_0\mp\beta \eta_\alpha$ for small $\eta_\alpha =2\epsilon\cot\alpha$ we obtain: 
\begin{equation}
\Gamma_{+}(t)\approx G(t)\big\{J_0 -\epsilon J_1\big\},\quad\quad\quad \Gamma_{-}(t)\approx-G(t)\big\{J_0 +\epsilon J_1\big\}, 
\end{equation}
with $\epsilon(t)$ and $G(t)$ given by (\ref{eqn:scale}) and $J_1=2\beta\cot\alpha\geq 0$. Each expression is exactly the first two terms of the expansions used by Pullin \& Wang [\onlinecite{Pullin:04b}], who, with a single point vortex approximation, obtained a relation for $2\beta$ as a function of the acceleration exponent $m$ (recall $U=Bt^m$). 
By computing $\beta$ from the slopes of $J_\pm$ at $\eta_\alpha=0$ for the range $0\leq m\leq 1$, we find that the expression for $J_1$ given in Ref.~[\onlinecite{Pullin:04b}] (their equation (3.27)) under-predicts by a nearly constant amount of $\Delta J_1 \approx 0.43\cot\alpha$, which corresponds to about 24\%. This offset is due to the error associated with the single point vortex approximation as compared to the vortex sheet solution used here. 

The major advantage of our model is that higher-order dependence on $\eta_\alpha$ comes built-in with the similarity solutions $J_\pm$ and $\omega_\pm$ since the parameter $\eta_\alpha$ appears explicitly in the governing equations (\ref{eqn:GE}). For example, as seen in figure~\ref{fig:Jcore}(\textit{a}), a reversal in the growth trend of the TE circulation $J_{+}$ occurs for $\eta_\alpha$ above about 0.75. This non-dimensional parameter is a similarity variable that `collapses' the effects of angle of attack and the time growth of the vortex spirals. It is akin to the similarity parameter for the Blasius boundary layer that collapses the viscous layer growth with downstream distance. Since $\eta_\alpha=2\epsilon(t)\cot\alpha$, then increasing time tracks with an increase of $\eta_\alpha$, while increasing $\alpha$ corresponds to a decrease of $\eta_\alpha$. Also shown in figure~\ref{fig:Jcore}(\textit{a}) is the non-dimensional bound circulation on the plate, which is given by $J_b=J_{-}-J_{+}$ since $J_o=0$. The dimensional bound circulation is $\Gamma_b=\big|\Gamma_{-}\big|-\Gamma_{+}>0$ and so is in opposite sense to that of the leading-edge sheet. This quantity is one indicator of the asymmetry in the total vortex structure. 

\begin{figure}
\begin{center}
\begin{minipage}{0.9\linewidth}\begin{center}
\includegraphics[width=0.99\textwidth, angle=0]{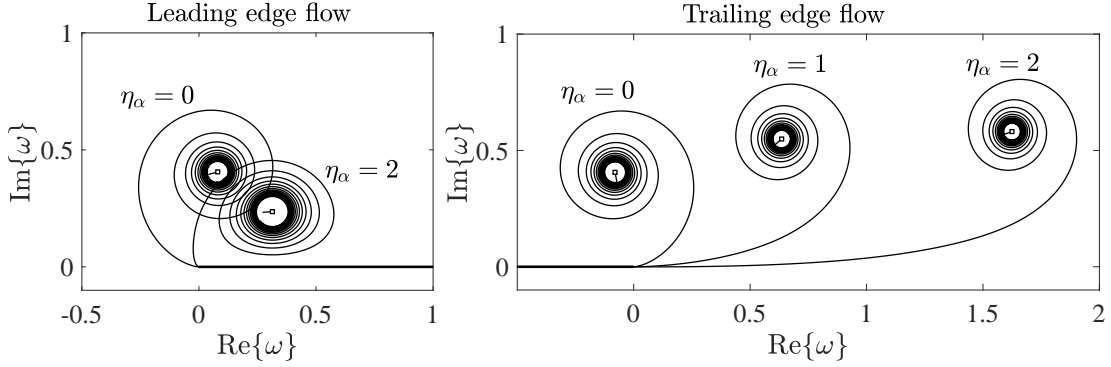}\\ \vspace{-10pt}(\textit{a}) Impulsive acceleration: $m=0$
\end{center}\end{minipage}
\\
\vspace{20pt}
\begin{minipage}{0.9\linewidth}\begin{center}
\includegraphics[width=0.99\textwidth, angle=0]{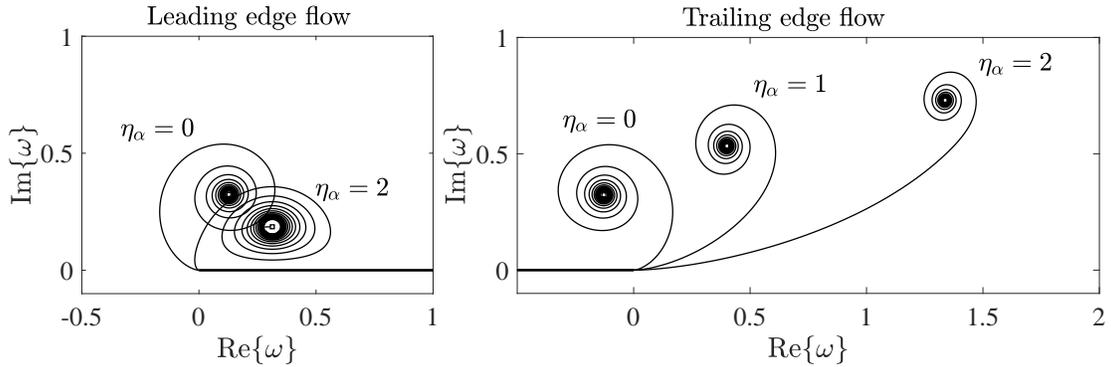}\\ \vspace{-10pt}(\textit{b}) Constant acceleration: $m=1$
\end{center}\end{minipage}
\caption{Sheet shapes in similarity space for flow near the leading edge, $-\cc{\omega}_{-}$ (left plots), and flow near the trailing edge, $\omega_{+}$ (right plots). Solutions with $J_o=0$ and different values of the parameter $\eta_\alpha=2\epsilon\cot\alpha$ are plotted as labeled, where $\eta_\alpha=0$ is the baseline with no effect of asymmetry. The sweeping motion of the free-stream represented by $\eta_\alpha$ is from left-to-right. (\textit{a})~Impulsive acceleration: $m=0$, and (\textit{b}) constant acceleration: $m=1$.}
\label{fig:mk}
\end{center}
\end{figure}
The asymmetric effects are best observed in the LE and TE sheet shapes. Some examples are plotted in figure~(\ref{fig:mk}) for different values of $\eta_\alpha$ and acceleration exponents $m$. The downstream convection of the TE spiral and the pronounced ellipticity of the LE spiral are familiar features. As the LE sheet grows in strength (recall figure~\ref{fig:Jcore}\textit{a}), the spiral core is pinned closer to the plate surface representing the coveted `LEV' known to increase the lift [\onlinecite{EllingtonCP:84a}] (so long as it remains attached). The LE flow separation will be discussed further in \S\ref{sec:lesp}. To give a better idea of the `rate' at which the LE and TE spiral cores convect downstream, figure~\ref{fig:Jcore}(\textit{b}) plots their positions as a function of $\eta_\alpha$ for the $m=0$ case.

The second set of similarity solutions we explored was defined by variable $J_o$ and $\eta_\alpha=0$, the latter corresponding to an infinitely small time $t=0^{+}$ after the onset of motion, or a plate moving normal to its chord, $\alpha=90^o$. Here, the effect of $J_o>0$ at the trailing edge is the same as $J_o<0$ at the leading edge, and vice versa. By (\ref{eqn:Kutta}), we see that the flow due to $J_o$ combines with the attached flow around the edge in either a constructive or destructive manner. In the former case, the sheet circulation increased linearly with $J_o$. The sheet shape remained unchanged, but its size enlarged such that the spiral core position moved along a constant ray drawn from the plate edge. For the latter case, the opposite is true: the circulation decreased, becoming zero at $|J_o|=1$, i.e. an exact canceling of the attached flow, and the sheet shrank to the point at the edge. This situation corresponds to the steady-state lift of a Joukowski airfoil satisfying the Kutta condition at its cusped TE ($\Gamma_{+}=0$) and with separation at the LE ignored ($\Gamma_{-}=0$). Namely, no vortex sheets are shed, but the bound circulation of $\Gamma_b=\Gamma_o=\pi cU\sin\alpha$ is established and equal to that at infinity. For $|J_o|>1$ a sheet of the opposite sign circulation must form on the pressure side of the plate, since the $J_o$ flow overcomes the attached flow due to the translation of the plate. This type of solution would be applicable to flapping wings or hovering flight at high angles of attack where a stroke reversal occurs. 

\subsection{The vortex force}\label{sec:Fv}
In this section we perform some manipulations on the vortex force in order to obtain a more intuitive expression. Replacing $\Gamma$ with the non-dimensional circulation parameter $\lambda$ as the integration variable in (\ref{eqn:Fv}) allows the vortex force to be rewritten as:
\begin{eqnarray}\label{eqn:Fv1}
F_t^{(v)}+iF_n^{(v)} = i\rho\der{}{t}\left\{\sum_\pm\Gamma_\pm \left[c\int_0^1\frac{(\zeta_\pm-\zeta_\pm^i)}{c}\mdi{\lambda}\right]\right\}
\equiv  i\rho\der{}{t}\left\{\sum_\pm\Gamma_\pm\big[cI_\pm\big]\right\}
\end{eqnarray}
where $I_\pm\big(t,\eta_\alpha,J_o\big)$ are symbols for the non-dimensional integrals above, which depend on time explicitly as well as implicitly through the parameters $\eta_\alpha$ and $J_o$. The quantities $c I_\pm$ can be interpreted as the positions of point vortices with circulations $\Gamma_\pm$ that preserve the force contribution on the plate due to each vortex sheet. While we have gained higher-order effects of asymmetry by absorbing $\epsilon$ into the non-dimensional governing equation through $\eta_\alpha$, the physical positions $Z_\pm$ of the sheets given by (\ref{eqn:wp}) and (\ref{eqn:wn}) are still only $O(\epsilon^2)$ accurate. As such, our calculation of (\ref{eqn:Fv1}) should be of the same order. This is accomplished by substituting $Z_\pm$ into (\ref{eqn:Jtrans}) to obtain $\zeta_\pm$ and these expressions into $I_\pm$, and then expanding the result for small $\epsilon$. 

First, however, a brief comment regarding the calculation of the force contribution from the leading-edge sheet $Z_{-}$ is appropriate. Since (\ref{eqn:wn}) involves the negative conjugate of $\omega_{-}$, then this operation must also be applied to the normal vector of the sheet to keep the force components consistent with the coordinate system of the problem. Also, the minus sign reflecting $\Gamma_{-} <0$ is canceled (i.e. made positive) by the opposite direction of integration along the leading-edge sheet as compared to the trailing-edge sheet. In effect, the force from $Z_{-}$ can be calculated in the same way as that from $Z_{+}$ and then applying the negative conjugate operation to the result. 

Returning to our task, expanding (\ref{eqn:Fv1}) and dropping terms $O(\epsilon^3)$ and higher yields:
\begin{subequations}
\begin{gather}
F_t^{(v)}+iF_n^{(v)} = \rho\der{}{t}\left\{c\epsilon G\Big[\epsilon\big(J_{-}T_{-}-J_{+}T_{+}\big)+i\big(J_{+}N_{+}+J_{-}N_{-}\big)\Big]\right\} \label{eqn:Fv2} \\
N_\pm = \int_0^1\mathrm{Re}\big\{\sqrt{\omega_\pm}\big\} \mdi{\lambda}, \quad\quad\quad
T_\pm = \int_0^1\mathrm{Im}\big\{\omega_\pm\big\}\mdi{\lambda}
\end{gather}
\end{subequations}
where $N_\pm(\eta_\alpha,J_o)$ will give rise to $O(1)$ plate-normal forces and $T_\pm(\eta_\alpha,J_o)$ to $O(\epsilon)$ plate-tangential forces. If the expansions used by Pullin \& Wang [\onlinecite{Pullin:04b}] are substituted it is found that $(J_{-}T_{-}-J_{+}T_{+})=0$ and so the remaining force is normal to the plate. However, as explained in \S\ref{sec:impulse}, this must necessarily be the case since the plate only delivers a net normal impulse to the fluid when $\Gamma_o=0$. One will also find that $(J_{+}N_{+}+J_{-}N_{-})=2J_0N_0$, where $J_0$ and $N_0$ are the values corresponding to the singular-order problem of Pullin [\onlinecite{Pullin:78a}], and thus we also recover their result that higher-order corrections to the (normal) vortex force vanish by cancellation. The cancellation occurs due to `mirror symmetries' inherent to the construction of their expansions. For our model this does not happen because the asymmetry is built-in at the level of the differential equation and is propagated through to $\omega_\pm$ and $J_\pm$. Note that so long as the sheets remain above the plane of the plate, then $N_\pm >0$ and $T_\pm >0$. In this case, the LE and TE structures work together to produce the normal force, whereas they work against each other in establishing any tangential force.

Now, in distributing the time derivative through each term of (\ref{eqn:Fv2}) we must account for both the explicit time dependence of $G(t)$ and $\epsilon(t)$, as well as the implicit time dependence of $J_\pm$, $N_\pm$ and $T_\pm$ via their dependence on $\eta_\alpha(t)$ and $J_o(t)$. This involves several applications of chain and product rule differentiation and results in a cumbersome expression. To maintain notational clarity we introduce an operator $\mathcal{F}_p(\mathcal{X})$ with argument $\mathcal{X}$ and parameter $p$ that is defined by:
\begin{eqnarray}\label{eqn:oper}
\mathcal{F}_p(\mathcal{X})\equiv \left[\left(Q+\frac{1}{p}\right)\mathcal{X}+\frac{1}{2}\eta_\alpha\der{\mathcal{X}}{\eta_\alpha}\right] + J_o\der{\mathcal{X}}{J_o}\left[\frac{R_v}{\dot{R}_v}\left(\frac{\dot{\Gamma}_o}{\Gamma_o}+\frac{\dot{U}}{U}\right)\right]. 
\end{eqnarray}
A dot indicates time differentiation and note that $\mathcal{F}_p$ is a linear operator. Also, the second grouping of terms vanishes entirely if either $\Gamma_o=0$ or $J_o$ is constant; it is assumed that $U\neq 0$ for $t>0$. The vortex force in (\ref{eqn:Fv2}) can then be expressed as:
\begin{eqnarray}\label{eqn:Fv3}
F_t^{(v)}+iF_n^{(v)} = \rho c\epsilon\dot{G}Q^{-1}\Big\{\epsilon\mathcal{F}_1\big(J_{-}T_{-}-J_{+}T_{+}\big)+i\mathcal{F}_2\big(J_{+}N_{+}+J_{-}N_{-}\big)\Big\}.
\end{eqnarray}
Again, $Q=(4m+1)/(2m+2)$ with $m$ as the acceleration exponent in the velocity $U(t)=Bt^m$. Moreover, using (\ref{eqn:scale}) we have:
\begin{equation}\label{eqn:dynP}
\rho c\epsilon\dot{G}Q^{-1}=\frac{1}{2}\rho U^2c\left(\frac{\sin^{5/3}\alpha}{\left[C_p\left(\frac{Ut}{c}\right)\right]^{1/3}}\right),
\end{equation}
thus providing a convenient form suitable to typical definitions of non-dimensional force coefficients using the dynamic pressure and chord length. Note that $Ut/c=(1+m)(s/c)$ where $s(t)$ is the distance traveled by the plate at time $t$ and so $s/c$ is the number of chords traveled. The less intuitive force scaling given in Ref. [\onlinecite{Pullin:04b}] can be recovered by substituting for $U=Bt^m$ and introducing their definitions of $K=C_p^{2/3}$ and $a=c^{1/2}B\sin\alpha$.


In the following section we will consider actual time-dependent flows by converting the similarity solutions to physical space and compare the results to existing computations. The main focus will be the forces exerted on the plate.

\section{Applied results}\label{sec:results}
This section presents time-dependent results of our model that are converted from the similarity space to the physical space via the transformation procedure depicted in figure~\ref{fig:chart}. In order to validate and exhibit the capabilities and limitations of the current inviscid model, we make comparison to different viscous simulations. The net circulation $\Gamma_o$ is set to zero and so $J_o=0$ as well.

For a given $m$ and $\eta_\alpha$-$J_o$ curve in the similarity space, the operators $\mathcal{F}_1$ and $\mathcal{F}_2$ in the vortex force (\ref{eqn:Fv3}) can be evaluated without any further input from the dimensional problem. For all computed cases we found that $\mathcal{F}_1\big(J_{-}T_{-}-J_{+}T_{+}\big)\ll \mathcal{F}_2\big(J_{+}N_{+}+J_{-}N_{-}\big)$, by at least three orders of magnitude, while $\mathcal{F}_1(J_\pm T_\pm)$ and $\mathcal{F}_2(J_\pm N_\pm)$ are individually all of the same order (recall $\mathcal{F}_p$ is linear). This validates the statement made in \S\ref{sec:impulse} that the tangential vortex force must be zero when $\Gamma_o=0$, since then the plate only delivers a net normal impulse to the fluid. The tangential vortex force is not computed to be precisely zero because the similarity solutions at the LE and TE are obtained independent of each other. As such, the near mutual cancellation is a good first indication that the composite solution is accurately capturing the full vortex dynamics.

\begin{figure}
\begin{center}
\begin{minipage}{0.32\linewidth}\begin{center}
\includegraphics[width=0.99\textwidth, angle=0]{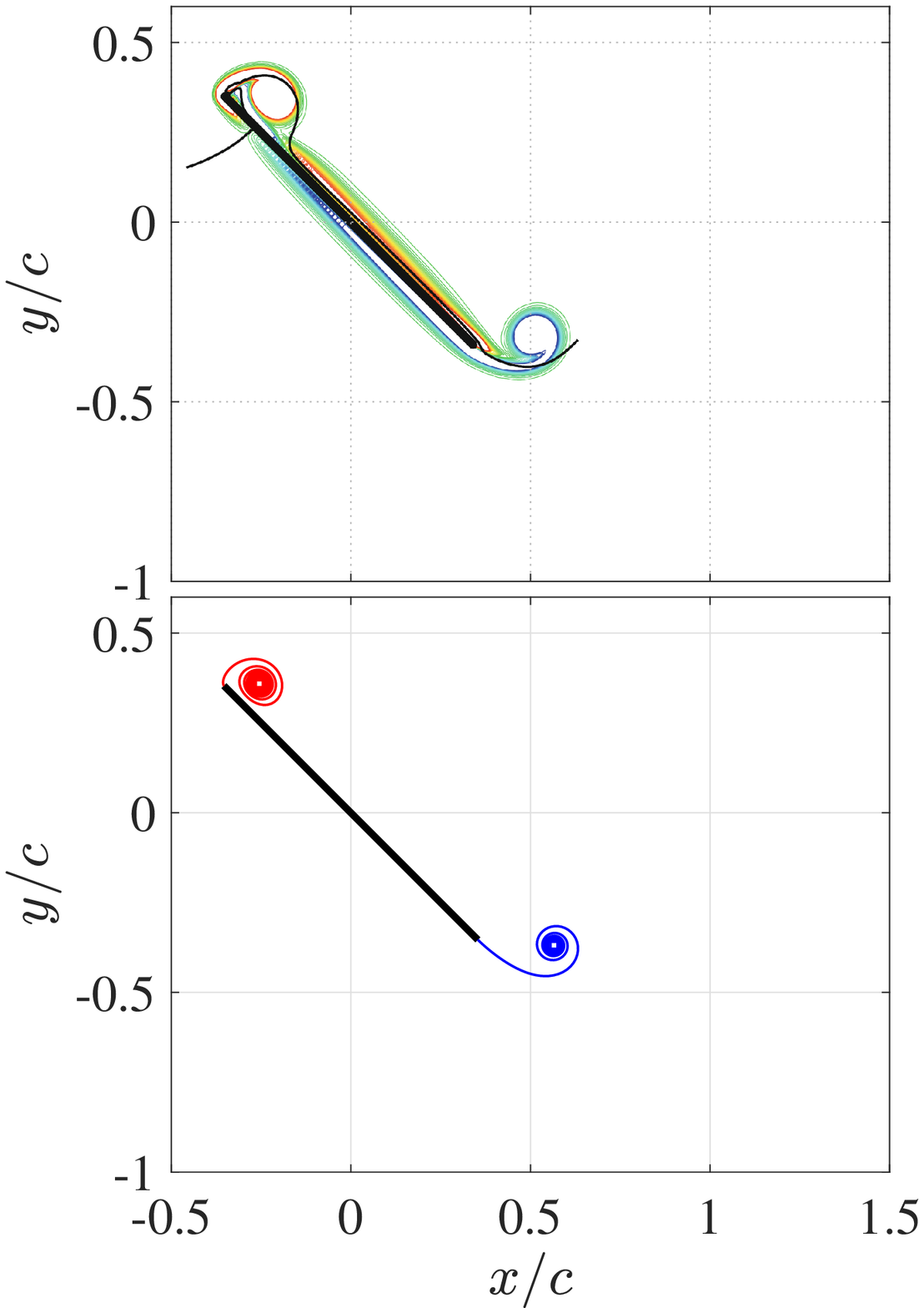}\\ (\textit{a}) $Ut/c=0.24$
\end{center}\end{minipage}
\begin{minipage}{0.32\linewidth}\begin{center}
\includegraphics[width=0.99\textwidth, angle=0]{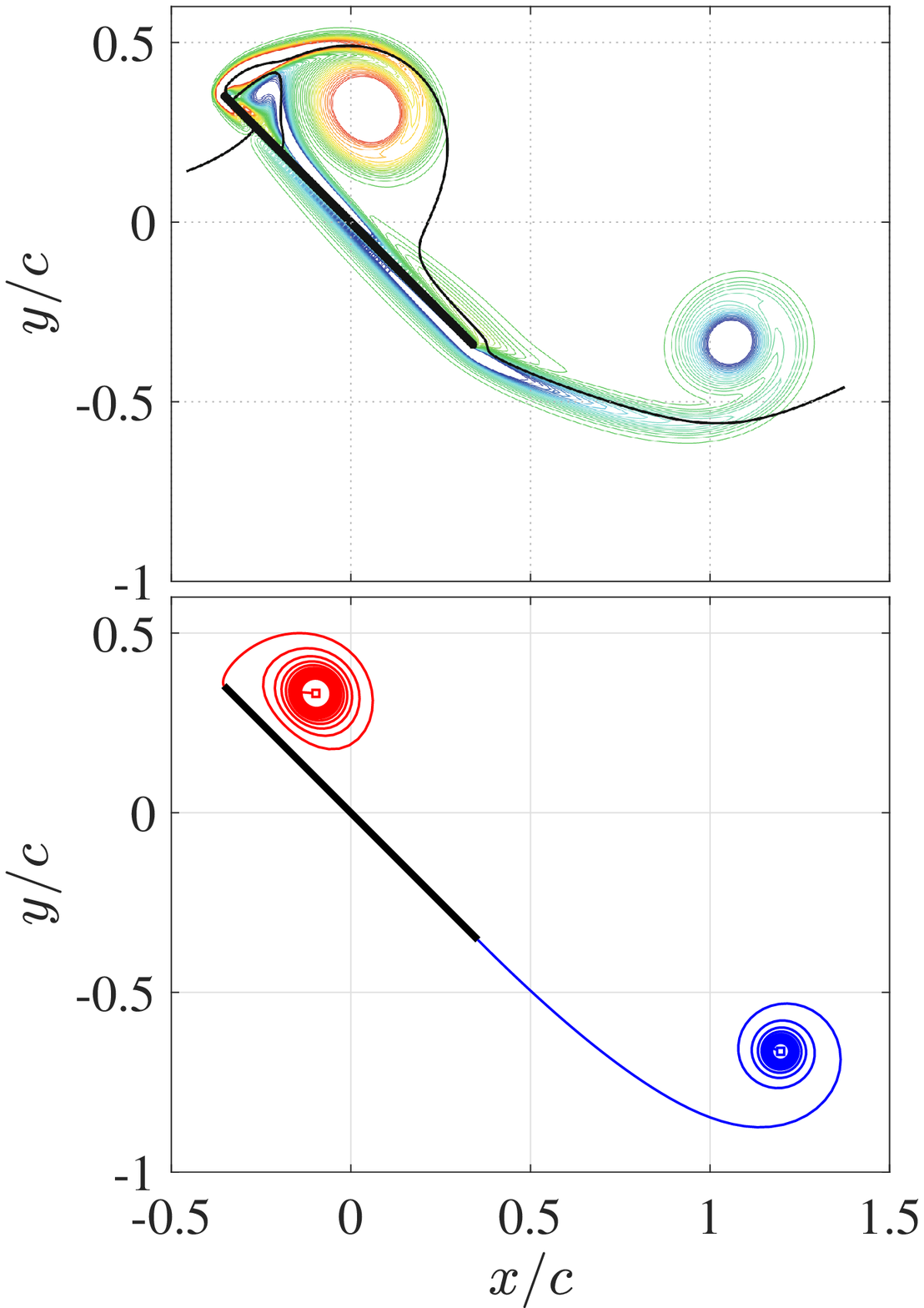}\\ (\textit{b}) $Ut/c=1$
\end{center}\end{minipage}
\begin{minipage}{0.32\linewidth}\begin{center}
\includegraphics[width=0.99\textwidth, angle=0]{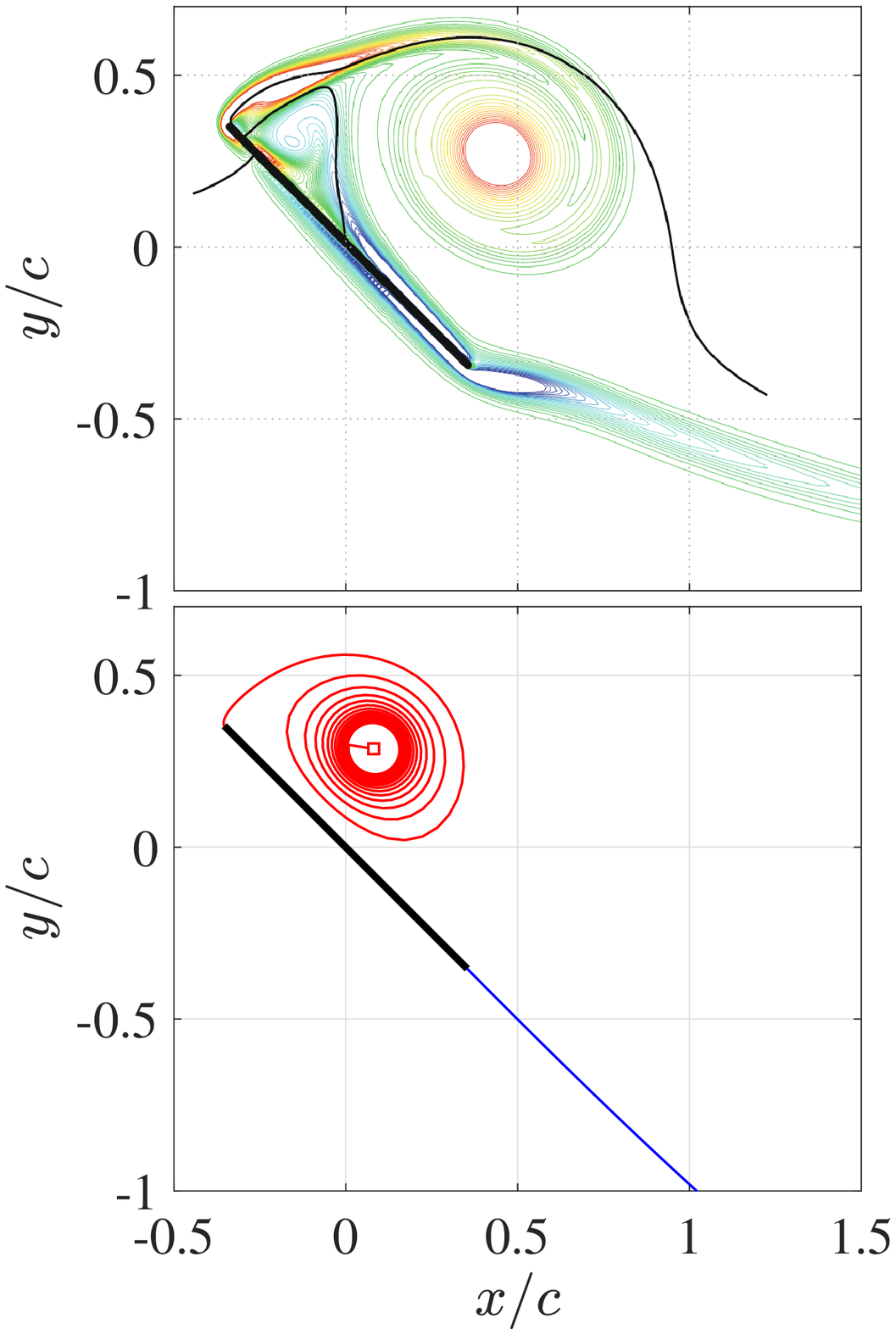}\\ (\textit{c}) $Ut/c=2.24$
\end{center}\end{minipage}
\caption{Flow structure comparison for the case of impulsive acceleration, $m=0$, at $\alpha=45^o$ at different convective times $Ut/c$ as labeled. (\textit{Top}) Vorticity contours from the viscous vortex particle method of Wang \& Eldredge [\onlinecite{EldredgeJD:13a}]. (\textit{Bottom}) Vortex sheet positions predicted by the present theory. Since $m=0$ then $Ut/c$ is equal to the number of chords traveled.}
\label{fig:Eld}
\end{center}
\end{figure}
\subsection{Impulsive acceleration}\label{sec:m0}
First, we make comparison to the viscous vortex particle method simulations (with $Re=1,000$) of Wang \& Eldredge [\onlinecite{EldredgeJD:13a}] for a flat plate that is impulsively accelerated, $m=0$, at $\alpha=45^o$. The top row of figure~\ref{fig:Eld} plots vorticity contours at several different `convective times' $Ut/c$; since $m=0$, then $Ut/c$ is also equal to the number of chords traveled. The bottom row of the figure plots the LE and TE vortex sheet positions from the current modeling of the same motion. There is good agreement early on with the size of the vortex spirals and core locations, at least up to one chord of travel. As expected, this begins to suffer as time increases, especially in the trailing-edge vortex sheet, which does not `see' the free-stream flow on the pressure side of the wing as the $y/c$ position of the spiral core dips below the trailing edge. However, the farther this vortex spiral is from the plate, the less influence its exact position will have on the force experienced by the plate.

\begin{figure}
\begin{center}
\begin{minipage}{0.49\linewidth}\begin{center}
\includegraphics[width=0.99\textwidth, angle=0]{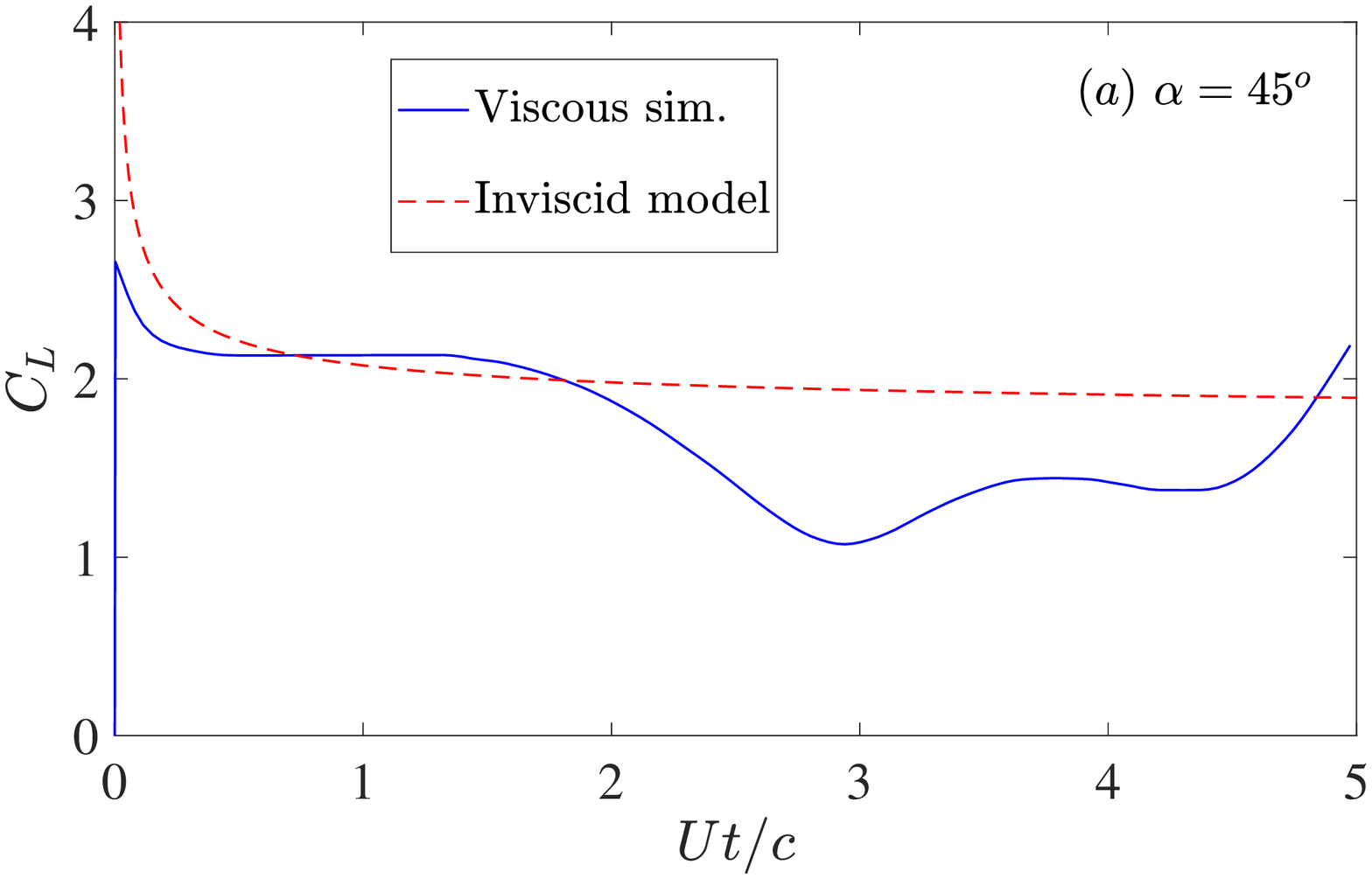}
\end{center}\end{minipage}
\begin{minipage}{0.49\linewidth}\begin{center}
\includegraphics[width=0.99\textwidth, angle=0]{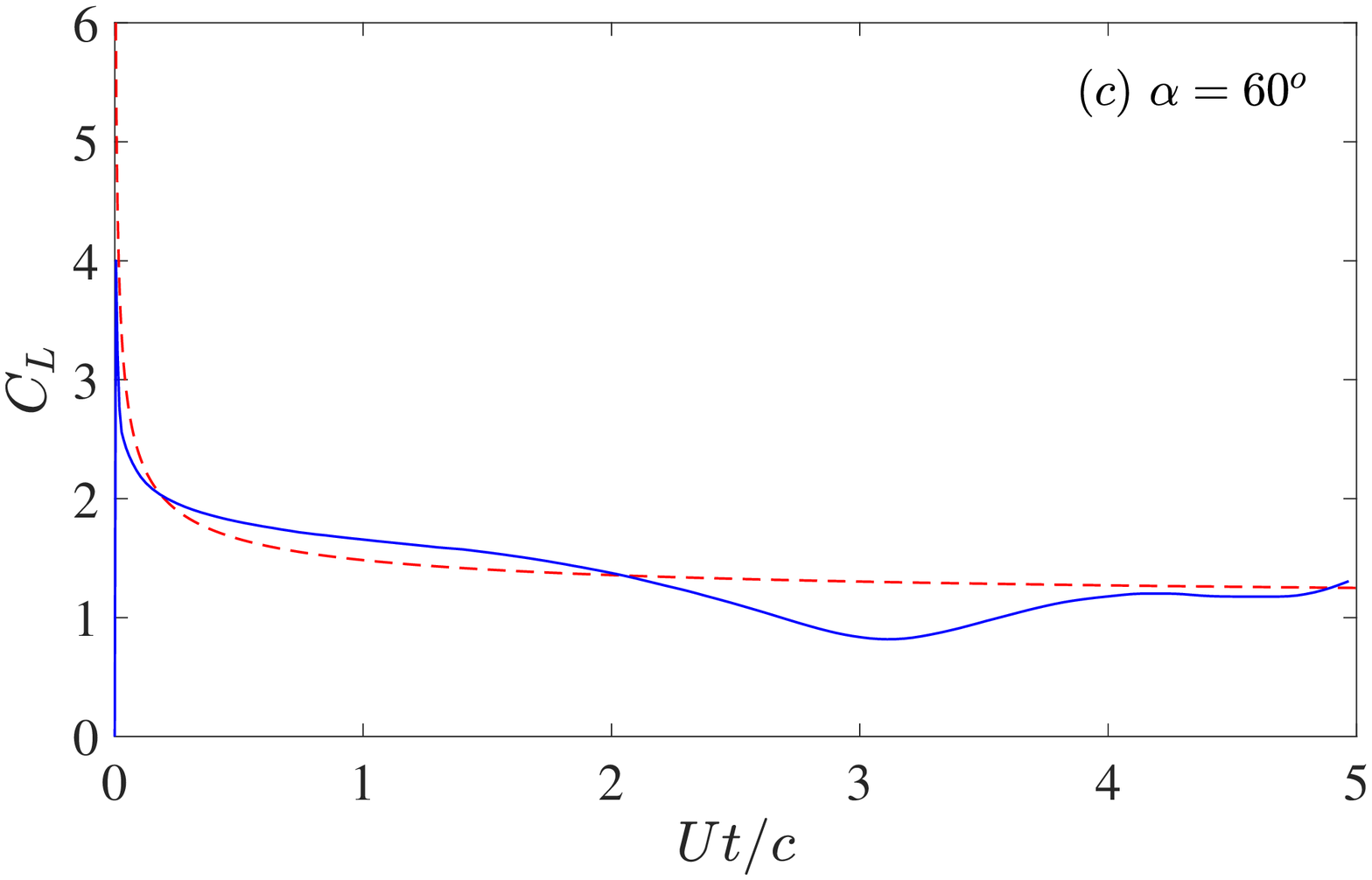}
\end{center}\end{minipage}
\\
\begin{minipage}{0.49\linewidth}\begin{center}
\includegraphics[width=0.99\textwidth, angle=0]{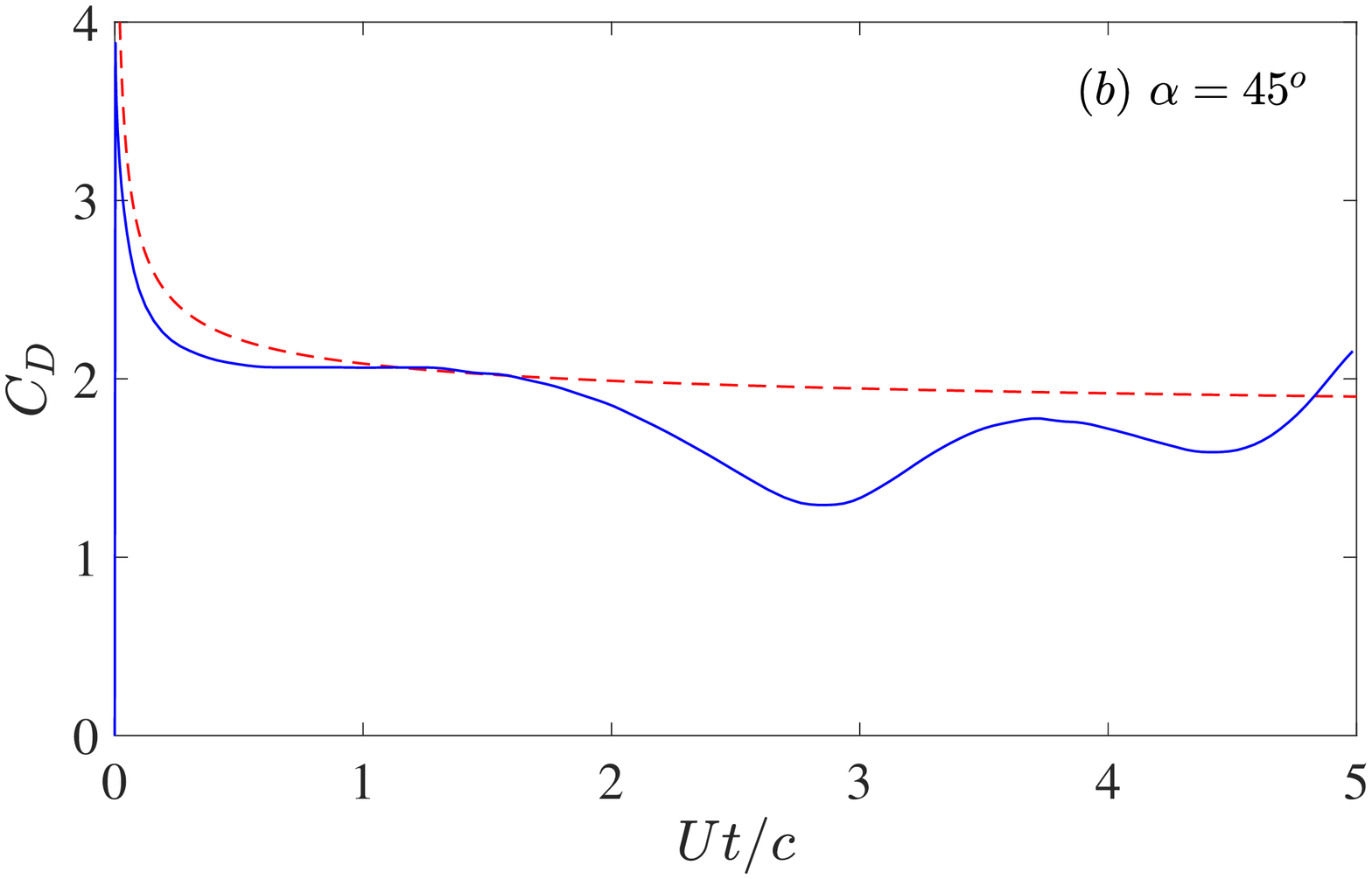}
\end{center}\end{minipage}
\begin{minipage}{0.49\linewidth}\begin{center}
\includegraphics[width=0.99\textwidth, angle=0]{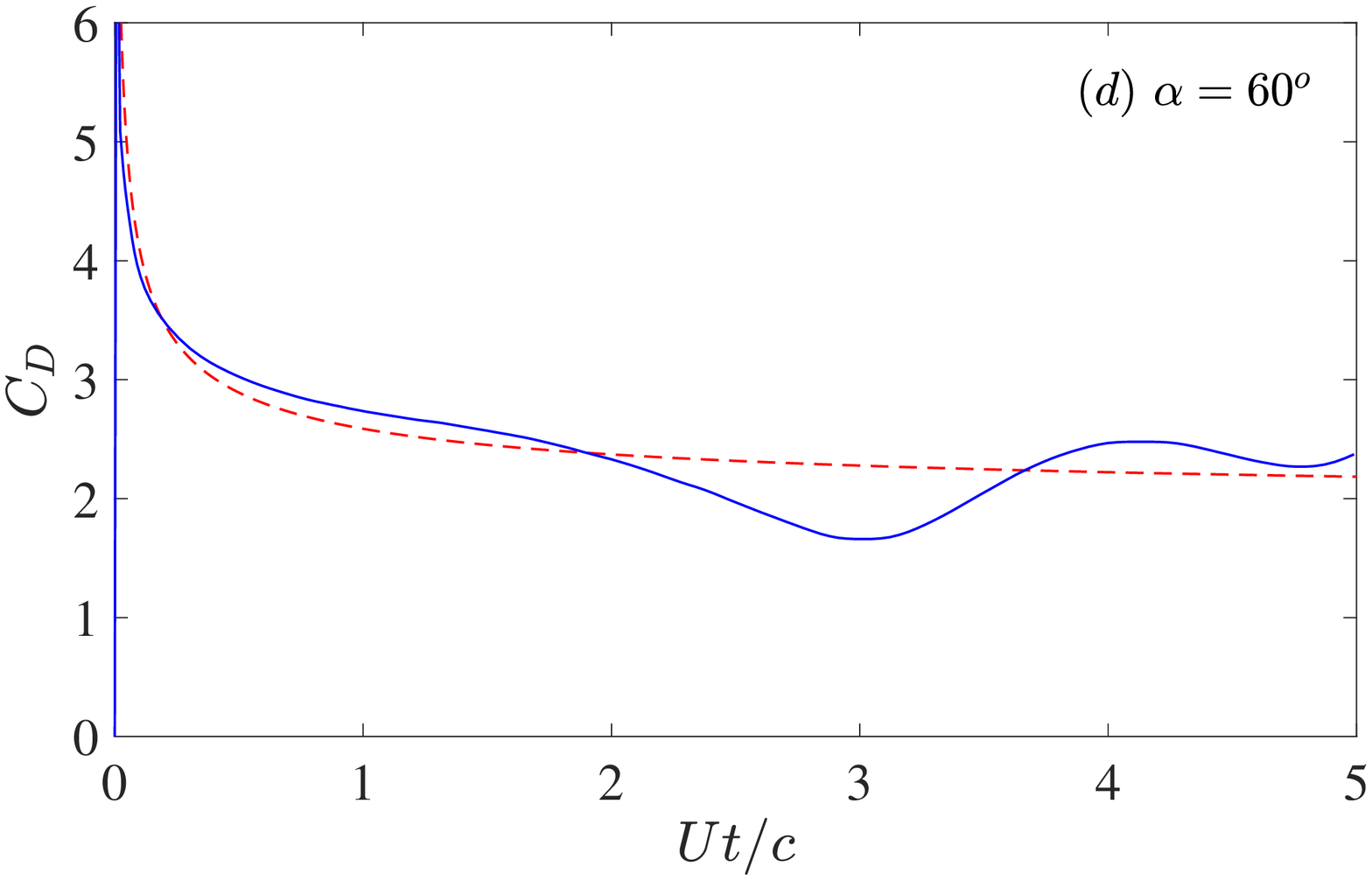}
\end{center}\end{minipage}
\caption{Comparison of lift and drag coefficients vs. $Ut/c$ for cases with impulsive acceleration, $m=0$, at two angles of attack. (\textit{a},\textit{b}) $C_L$, $C_D$ at $\alpha=45^o$; (\textit{c},\textit{d}) $C_L$, $C_D$ at $\alpha=60^o$. The Navier-Stokes simulations are from Ref. [\onlinecite{EldredgeJD:19a}]. The legend applies to all plots. Since $m=0$ then $Ut/c$ is equal to the number of chords traveled.}
\label{fig:Eld_f}
\end{center}
\end{figure}
To this end, a better metric of the model performance is given by comparison of lift and drag coefficients, $C_L$ and $C_D$. For this purpose, we use the results of the high-fidelity Navier-Stokes simulations (with $Re=500$) from Darakananda \& Eldredge [\onlinecite{EldredgeJD:19a}]. The force coefficients are shown in figure~\ref{fig:Eld_f} for two angles of attack: $\alpha=45^o$ and $60^o$. The inviscid model of the current paper provides a fair prediction up to $Ut/c\approx 2$ or about two chords of travel. Beyond this time, secondary vortex structures begin to form (see figures 4 and 11 of Ref. [\onlinecite{EldredgeJD:19a}]) and the unsteady forces will begin to oscillate due to the periodic shedding of LE and TE vortices. The inviscid model, being the composite of two self-similar flows at the edges, will not explicitly capture this physics as the flow developments at the edges can no longer be considered independent and thus self-similar. More will be said about this limitation of the model in the following subsection. 

\subsection{Constant acceleration}\label{sec:m1}
Next, we consider the case of constant acceleration, $m=1$, and compare with the Navier-Stokes simulations of Pullin \& Wang [\onlinecite{Pullin:04b}] for the flow around a thin elliptical airfoil; the elliptical cross-section has a minor-to-major axis ratio $e=0.125$. They provided numerical values of the physical parameters, but without specifying units: $c=2$, $B=4$. In effect, the time $t$ can be taken in seconds and the fluid to be of unit density such that the Reynolds number is $Re=Uc/\nu=800$ at $t=1$~s. We adhere to their specifications as well as their plotting of unscaled `dimensional' forces. 

\begin{figure}
\begin{center}
\begin{minipage}{0.49\linewidth}\begin{center}
\includegraphics[width=0.99\textwidth, angle=0]{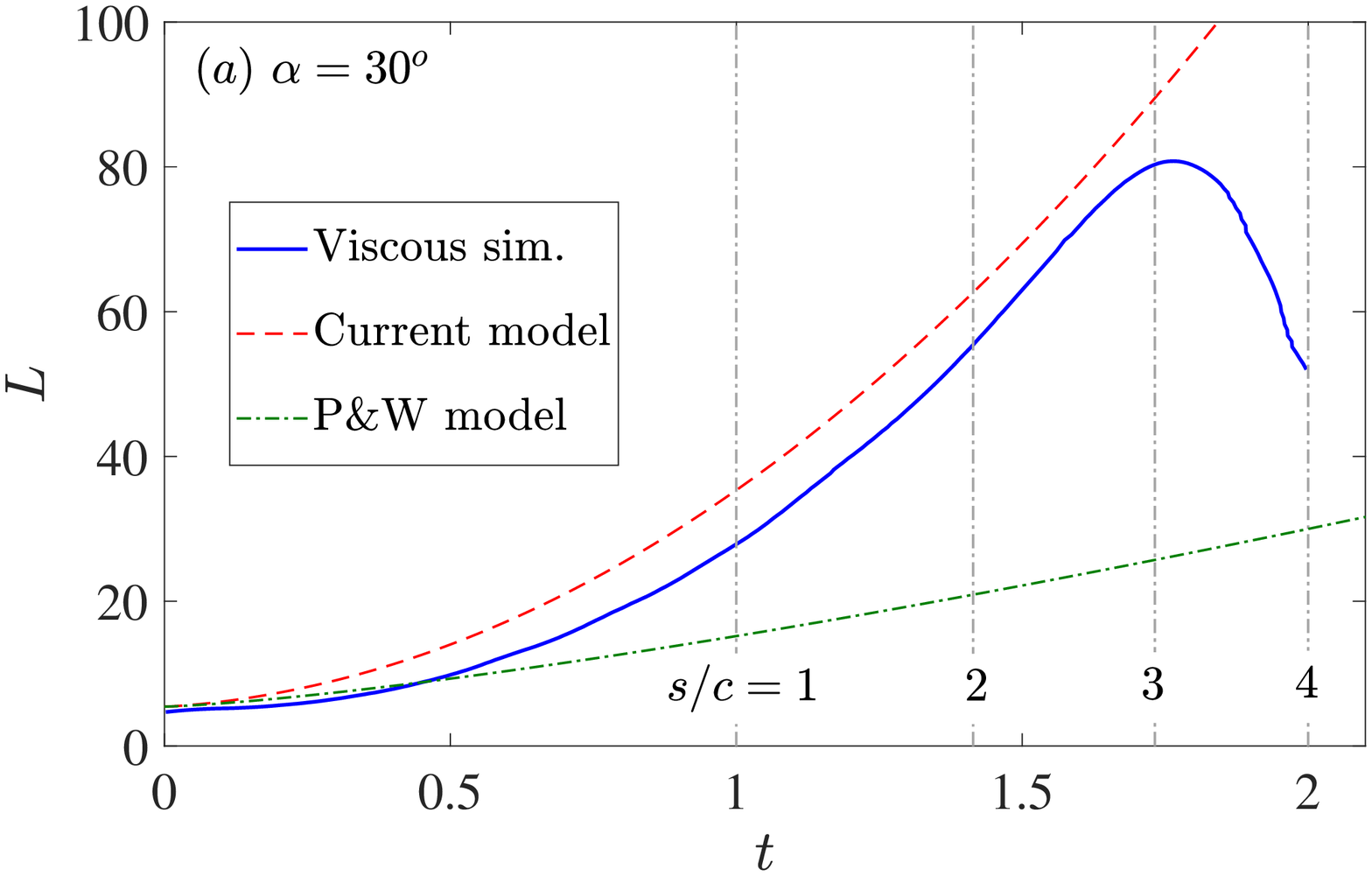}
\end{center}\end{minipage}
\begin{minipage}{0.49\linewidth}\begin{center}
\includegraphics[width=0.99\textwidth, angle=0]{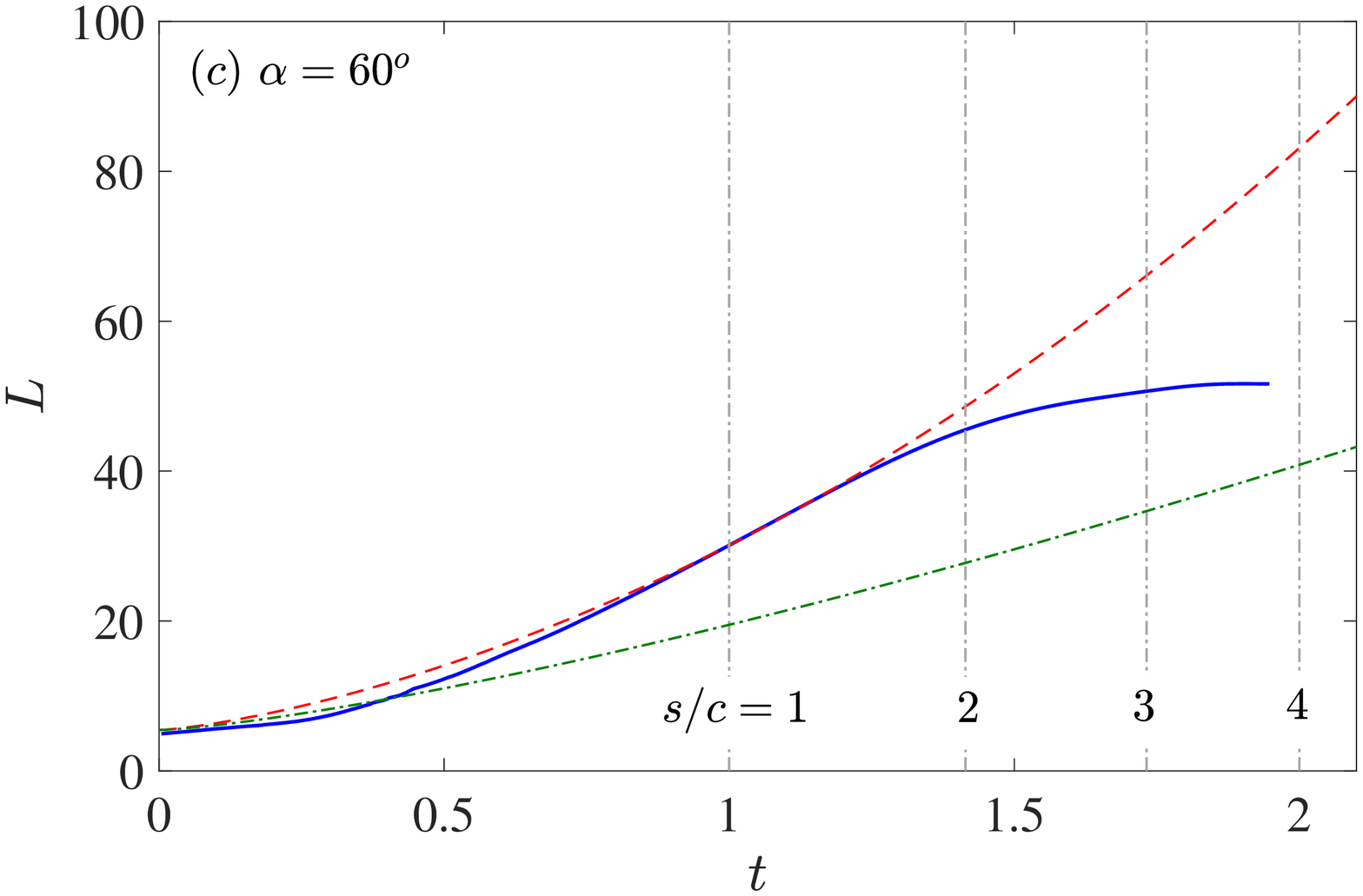}
\end{center}\end{minipage}
\\
\begin{minipage}{0.49\linewidth}\begin{center}
\includegraphics[width=0.99\textwidth, angle=0]{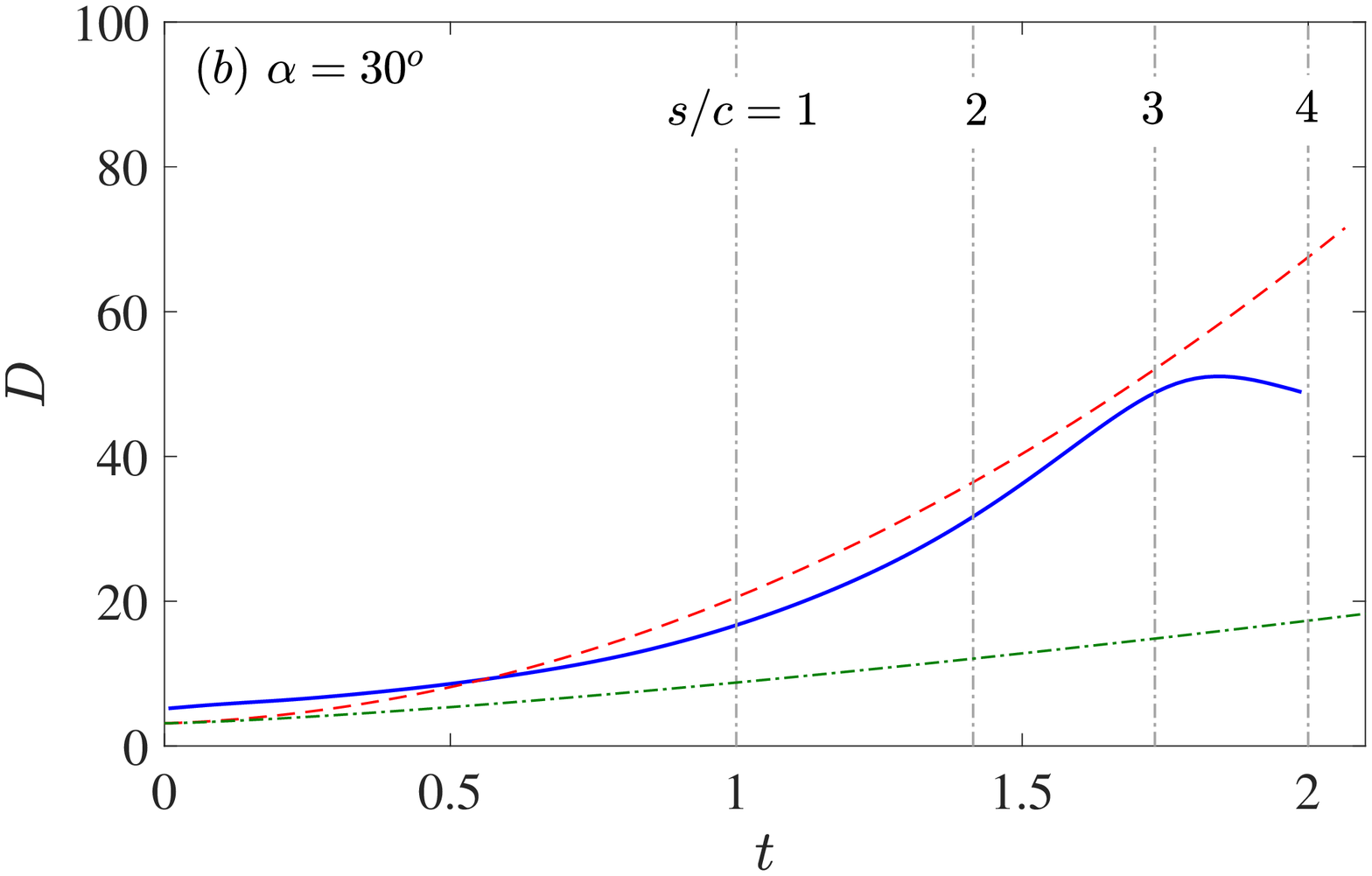}
\end{center}\end{minipage}
\begin{minipage}{0.49\linewidth}\begin{center}
\includegraphics[width=0.99\textwidth, angle=0]{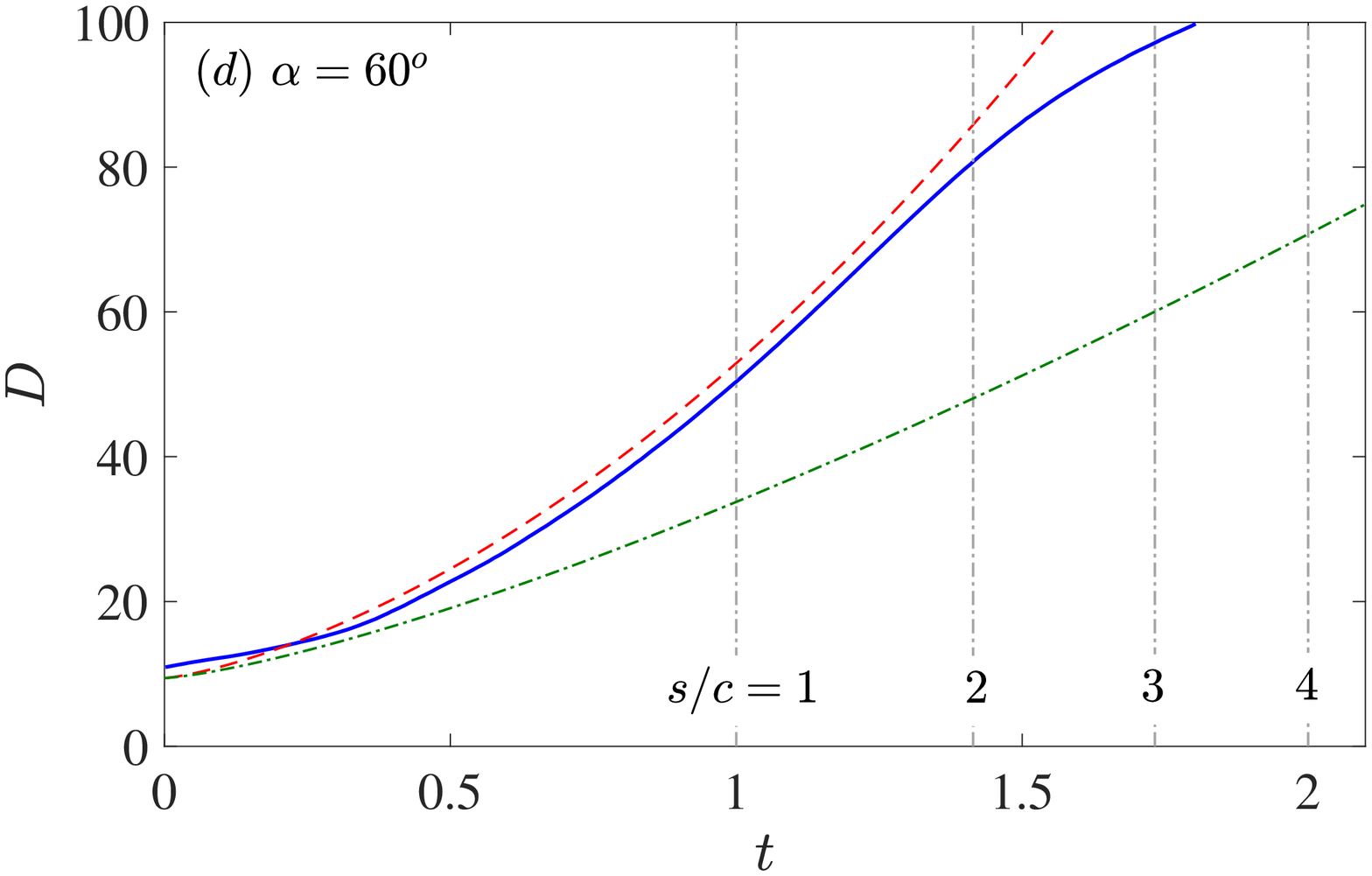}
\end{center}\end{minipage}
\caption{Comparison of lift and drag forces vs. $t$ for cases with constant acceleration, $m=1$, at two angles of attack. (\textit{a},\textit{b}) $L$, $D$ at $\alpha=30^o$; (\textit{c},\textit{d}) $L$, $D$, at $\alpha=60^o$. The Navier-Stokes simulations correspond to an ellipse of minor-to-major axis ratio $e= 0.125$ from Pullin \& Wang~[\onlinecite{Pullin:04b}]; their single vortex model is also shown and labeled as `P\&W model'. The legend applies to all plots. Vertical gray lines mark when the airfoil has traveled 1, 2, 3, and 4 chord lengths as labeled.}
\label{fig:PW_f}
\end{center}
\end{figure}
Figures~\ref{fig:PW_f}(\textit{a})-(\textit{d}) plot the lift, $L$, and drag, $D$, as functions of time $t$ from the viscous simulations of the elliptical airfoil for angles of attack $\alpha=30^o$ and $\alpha=60^o$. On each panel are vertical lines marking the times at which the airfoil has traveled 1, 2, 3, and 4 chord lengths. The forces exerted on the flat plate performing the same motions as modeled in the current paper are also shown and exhibit rather good agreement at least to two chord lengths of travel. For lower $\alpha$ the trend is followed to almost $s/c=3$ with some noticeable offset, while for higher $\alpha$ the quantitative match for $s/c<2$ is much closer. The inertial force on the elliptical wing is due to the non-zero area of the body and it is worth noting that that this is comparable to the \textit{constant} added-mass force of the zero-thickness flat plate, which is quantified by the force values at $t=0$ in the figure. As such, the vortex force does the overwhelming majority of the predictive work in the inviscid model. Also plotted is the prediction from the single point vortex model of Pullin \& Wang [\onlinecite{Pullin:04b}], which we recall contains no correction to the force due to asymmetry of the flow. We see that the current model provides a significant improvement in the prediction and can be attributed to the more effective representation of the vortex dynamics.

\begin{figure}
\begin{center}
\begin{minipage}{0.49\linewidth}\begin{center}
\includegraphics[width=0.99\textwidth, angle=0]{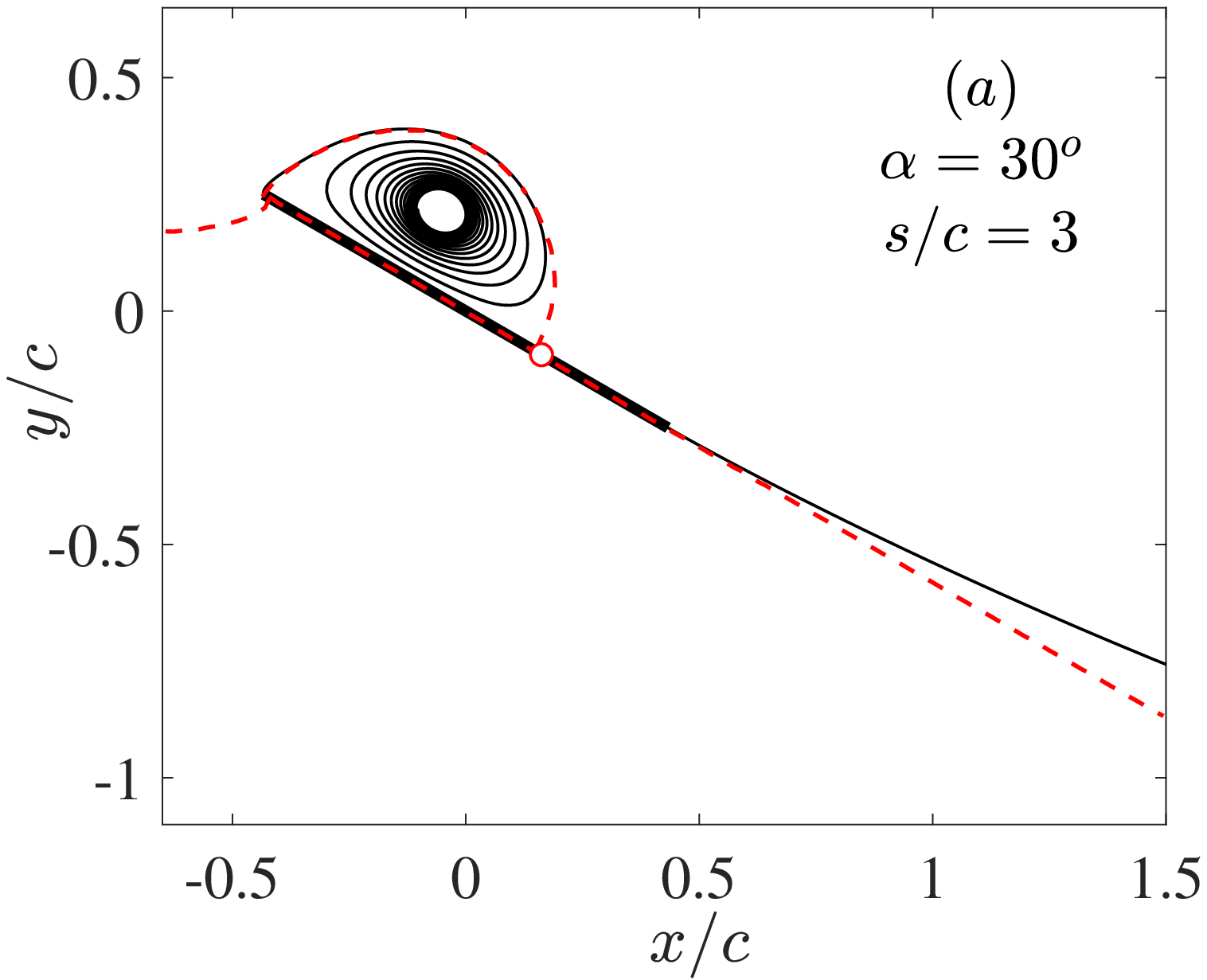}
\end{center}\end{minipage}
\begin{minipage}{0.49\linewidth}\begin{center}
\includegraphics[width=0.99\textwidth, angle=0]{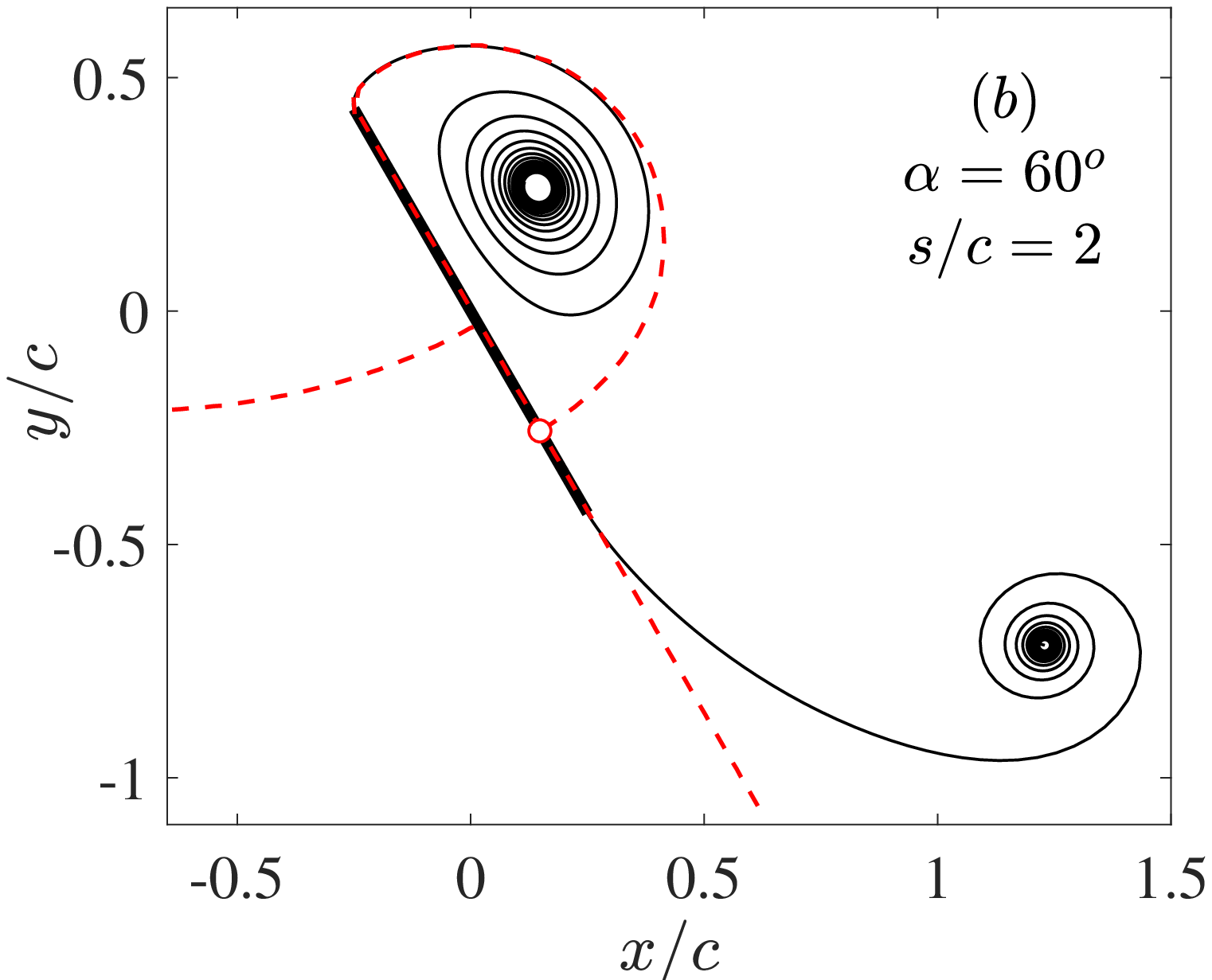}
\end{center}\end{minipage}
\caption{The modeled vortex structure at the time of near maximum lift for the cases shown in figure~\ref{fig:PW_f}. The stagnation streamline corresponding to the flow induced by just the leading-edge vortex sheet is also plotted (dashed red lines). The corresponding suction-side stagnation point is marked by the circle symbol. (\textit{a}) $\alpha=30^o$, chords traveled: $s/c=3$. The TE spiral is out of frame. (\textit{b}) $\alpha=60^o$, chords traveled: $s/c=2$.}
\label{fig:PW}
\end{center}
\end{figure}
We note that the major deviation of the model from the simulations in figure~\ref{fig:PW_f} begins at around the time of maximum lift. This corresponds to the first LE vortex shedding event. In other words, the vortex has grown so large that it can no longer remain attached to the plate and begins to convect downstream, thus giving a negative contribution to the lift. However, we can attempt to predict the initiation of this event, which we note is signaled by the arrival of the suction-side stagnation point, due to the reattached LE flow, at the trailing edge [\onlinecite{Mohseni:08l,RivalDE:14a,Mohseni:17c}]. Figures~\ref{fig:PW}(\textit{a}) and (\textit{b}) show the vortex sheet structures of the cases $\alpha=30^o$ and $60^o$ at the time of near maximum lift, which respectively occur at about $s/c=3$ and $s/c=2$ (figure~\ref{fig:PW_f}). At the times of maximum lift we find that $\epsilon^2\approx 1.1$ for each angle of attack case shown in figure~\ref{fig:PW}. Recalling that $\epsilon^2=R_v/c$ is a measure of the vortex spiral size relative to the plate chord, we might expect that $R_v\approx c$ is a good indication of the shedding event. Since the LE and TE flows are solved independent of each other, the composite stream function by superposition of the two flows is technically not valid; the streamlines do not have the visual character one would expect from a uniformly valid solution. However, at the times under consideration, the flow near the plate is dominated by the LE vortex sheet. Therefore, we estimate the suction-side stagnation point from the steam function of just this sheet. The stagnation streamlines are also plotted in figures~\ref{fig:PW}(\textit{a}) and (\textit{b}). While the estimated stagnation point has not quite yet reached the trailing edge, we note that inclusion of the effect of the flow induced by the TE vortex sheet will act to bring the point closer to the edge. From this and the metric $R_v\approx c$, we conclude that the model provides a reasonable prediction of the initial LE vortex shedding event.

The results of \S\ref{sec:m0} and \S\ref{sec:m1} indicate that the model is reasonably valid for a non-negligible distance/time traveled by the plate at different angles of attack and acceleration exponents. Hence, the conditions in (\ref{eqn:conds}) do not appear to be too strict. 

\begin{figure}
\begin{center}
\begin{minipage}{0.45\linewidth}\begin{center}
\includegraphics[width=0.99\textwidth, angle=0]{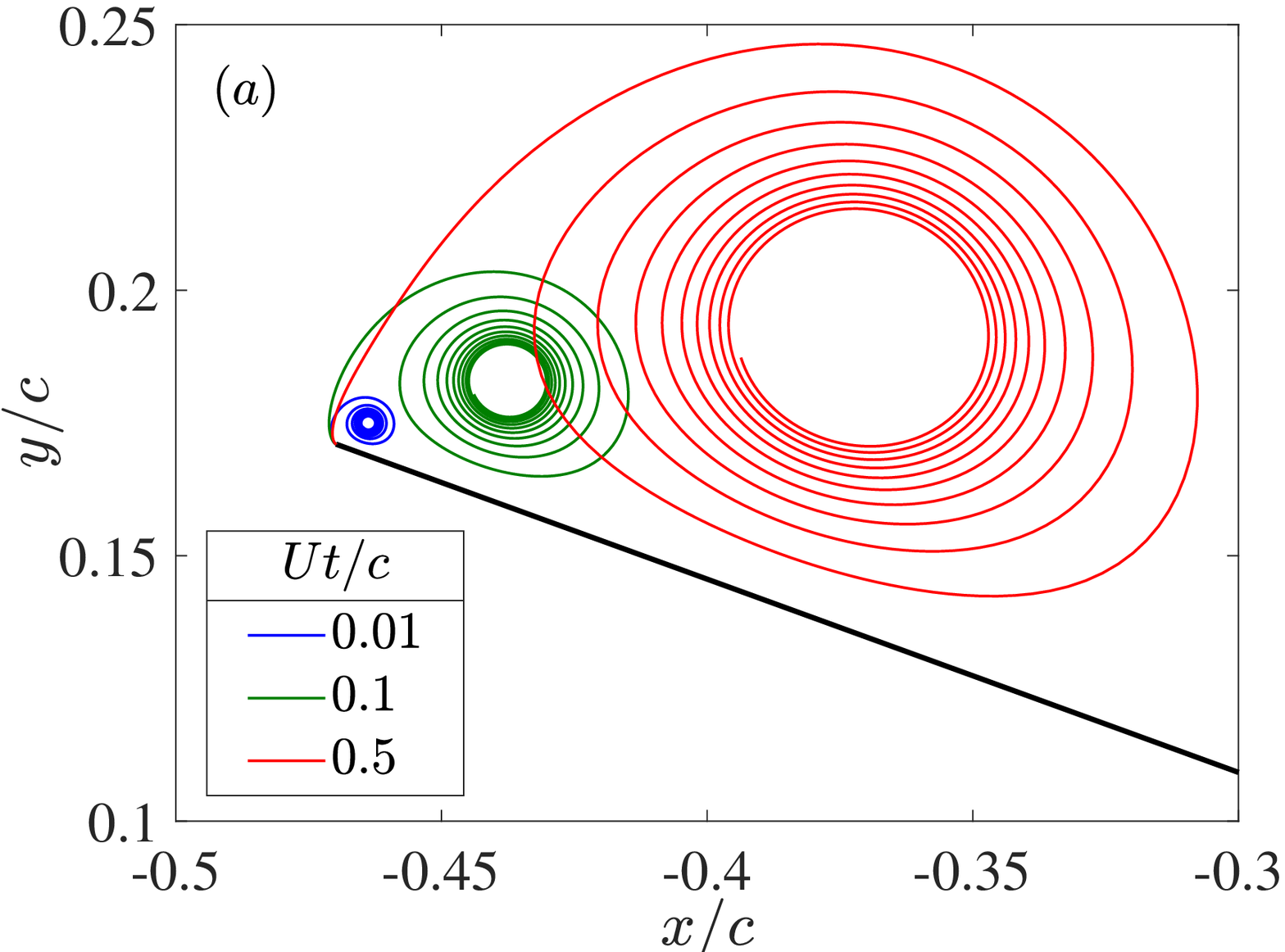}
\end{center}\end{minipage}
\begin{minipage}{0.45\linewidth}\begin{center}
\includegraphics[width=0.99\textwidth, angle=0]{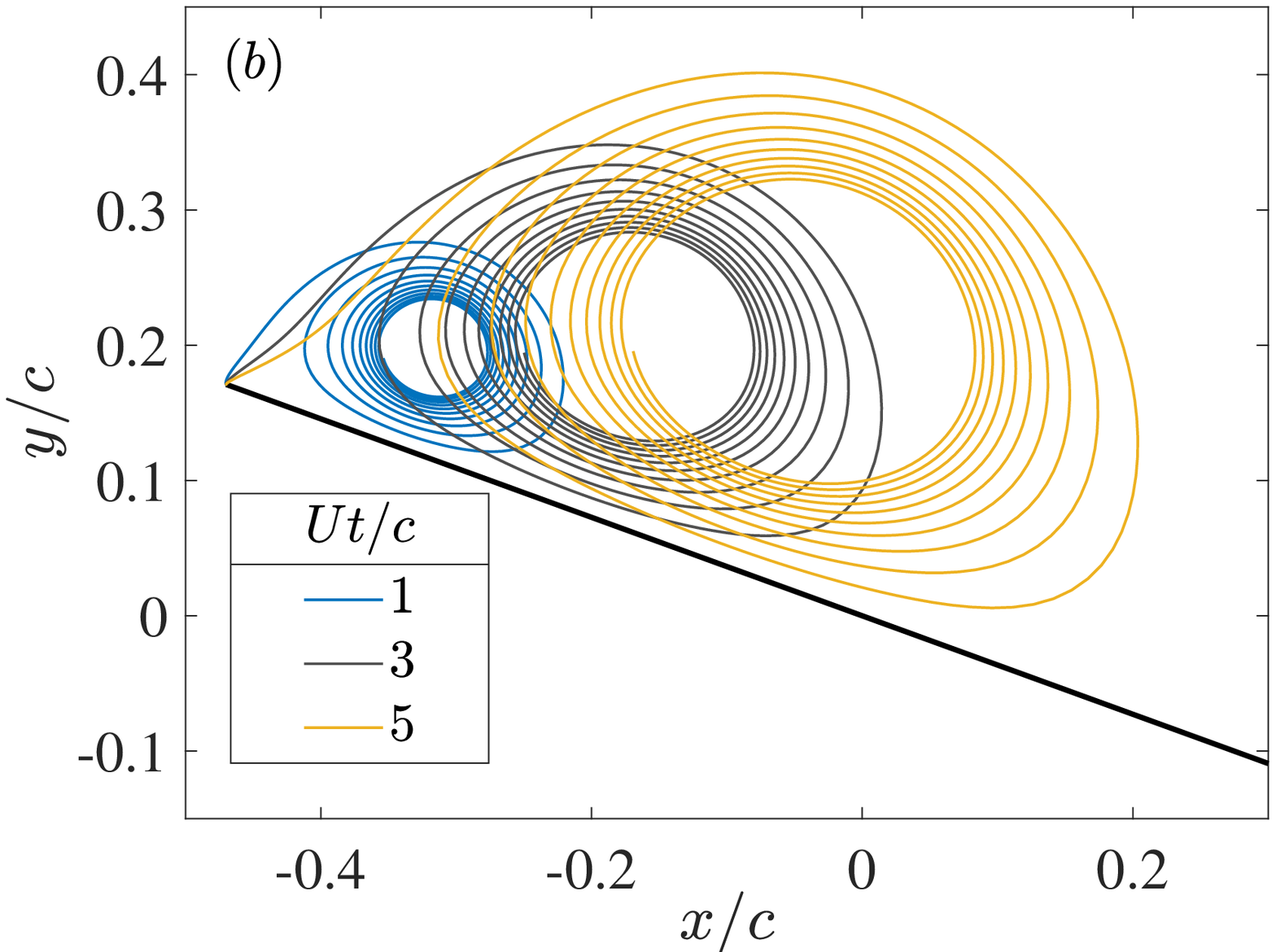}
\end{center}\end{minipage}
\begin{minipage}{0.47\linewidth}\begin{center}
\includegraphics[width=0.99\textwidth, angle=0]{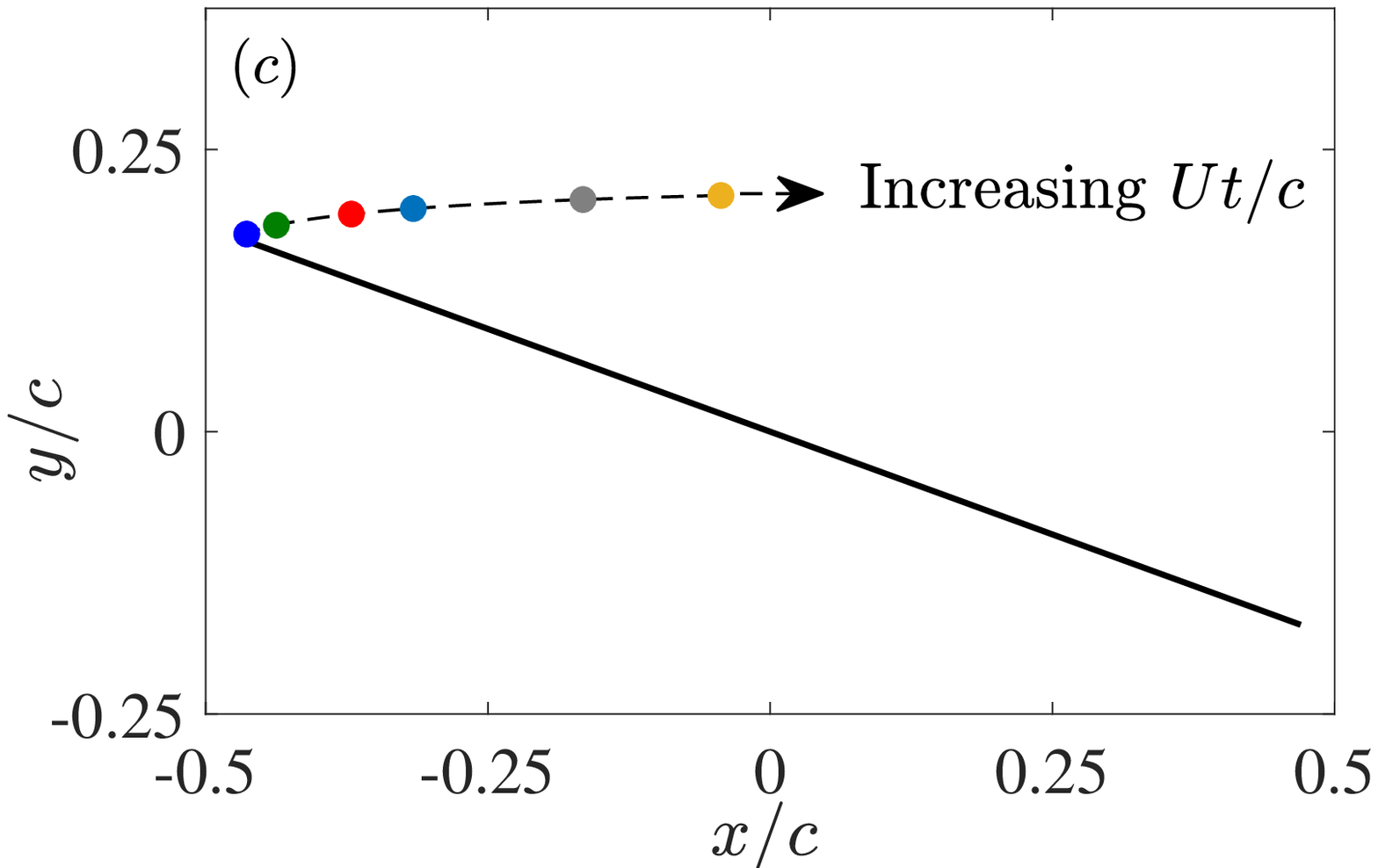}
\end{center}\end{minipage}
\begin{minipage}{0.47\linewidth}\begin{center}
\includegraphics[width=0.99\textwidth, angle=0]{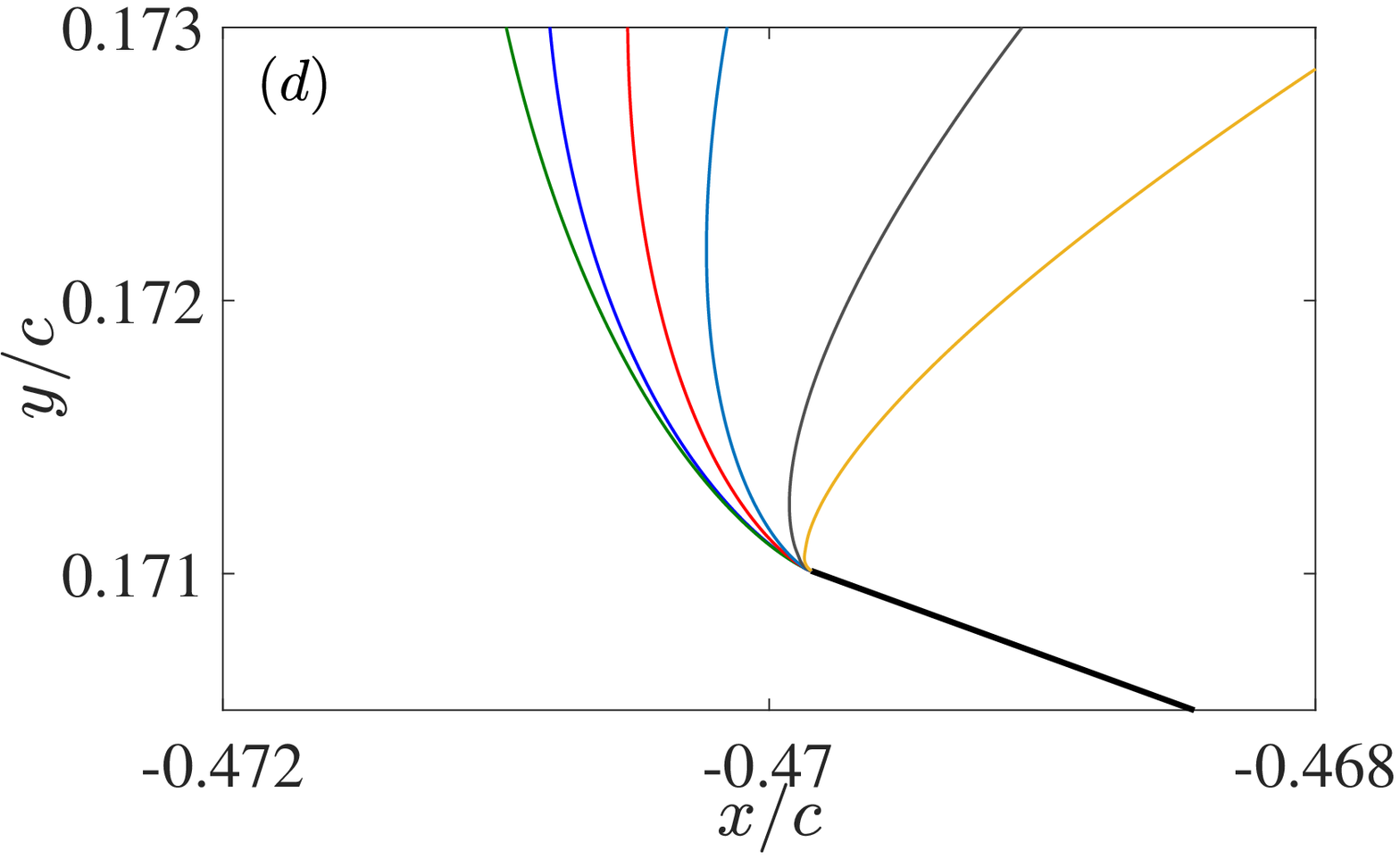}
\end{center}\end{minipage}
\caption{Sheet shapes in the physical plane for an impulsively accelerated plate at $\alpha=20^o$ for different convective time units $Ut/c$ as labeled. Note the difference in axis ranges between (\textit{a}) early times and (\textit{b}) later times. (\textit{c}) Trajectory of the spiral core location (dashed line) with dots color-coded to the $Ut/c$ values. (\textit{d}) Magnified view near the LE showing all cases.}
\label{fig:LE}
\end{center}
\end{figure}
\subsection{Separation at the leading edge}\label{sec:lesp}
Having sufficiently validated our model, we now turn our attention to the solution at the leading edge. While the shedding angle of the LE vortex sheet theoretically must be tangential to the plate [\onlinecite{Pullin:78a}], the region within which this occurs becomes increasingly smaller. This gives the appearance of a `kinked' sheet shape as can be seen in figure~\ref{fig:LE}, which plots the sheet shapes at several different convective times $Ut/c$ for the case of a plate with impulsive acceleration, $m=0$, at an angle of attack $\alpha=20^o$. Figure~\ref{fig:LE}(\textit{a}) plots early times when the model is valid. At $Ut/c=0.01$, which is a typical $\Delta t$ time-step in simulations, the rolled-up sheet effectively provides a small but rounded leading edge as expected. At slightly later times the sheet appears to bend almost normal to the plate. It is worth noting that at low-to-moderate angles of attack, the distortion of the sheet shape due to the asymmetric flow occurs rather early. For example, when $\alpha=20^o$ and $Ut/c\approx 0.2$, then the similarity variable has already achieved the value $\eta_\alpha\approx 2$; see figure~\ref{fig:mk}(\textit{a}) for the unscaled shapes in the similarity space and figure~\ref{fig:Jcore}(\textit{a}) for the non-dimensional circulation variation.

Figure~\ref{fig:LE}(\textit{b}) plots later times including some that are well beyond when the model is applicable (note the difference in axis ranges compared to plot (\textit{a})). By comparison, we see that the sheet begins to bend aft of the leading edge ($Ut/c=3$) and an inflection point where the curvature changes sign becomes increasingly evident ($Ut/c=5$). These latter two features are likely indicators of the LE vortex `pinch-off' event, whereby the vortex spiral is convected downstream by the sweeping free-stream flow and effectively severs its `umbilical' sheet to the leading edge. This tendency toward downstream convection is evident from the plot of the spiral core trajectory shown in figure~\ref{fig:LE}(\textit{c}). Whether or not the LE spiral in an actual flow ceases to accumulate an appreciable amount of shed circulation after the time when the sheet develops an inflection point requires further work, however.

Figure~\ref{fig:LE}(\textit{d}) shows a magnified view very near the leading edge for all of the cases in plots (\textit{a}) and (\textit{b}). This highlights the vanishingly small region in which the sheet must turn and become tangential to the plate chord. The resolution that would be required to capture this feature in a truly unsteady simulation would likely be prohibitive and advocates the practical necessity of a shedding criterion like the LESP. The present results could be used to inform the critical LESP value. An alternative is the vortex-entrainment sheet model of DeVoria \& Mohseni [\onlinecite{Mohseni:19d}], which naturally allows a non-tangential shedding angle when entrainment occurs at the edge. It is no surprise, then, that mass entrainment into a viscous shear layer would occur here and could be viewed as the `source' of the fluid contained within the closed region of a recirculating separated flow.

\section{Concluding remarks}
In this paper we considered the canonical problem of a flat-plate airfoil accelerating in an inviscid fluid at a constant angle of attack. The objective was to obtain a higher-order theoretical model with the separated flow represented as continuous vortex sheets. The full problem statement for the flat plate was approximated with self-similar solutions at both the leading and trailing edges. In contrast to previous works, we expanded the attached outer flow to higher order rather than the sheet positions and circulations. As such, the attached flow contains a regular part representing the induction of distant effects, as well as the usual singular part necessitating the physical flow to separate at the sharp edge. It was shown that the higher-order regular part of the outer flow expansion corresponds to the sweeping component of the free-stream flow parallel to the plate. Moreover, we introduced a similarity variable that collapses the temporal growth of the vortex spiral length scale with the effect of angle of attack. Through this parameter the sweeping flow is brought into the same order as the singular flow. In other words, the effect of asymmetry is built-in at the level of the governing equation which, in turn, is only trivially altered from the original equation that determines the singular-order flow alone.

Using this new self-similar model, we constructed composite flow solutions in the physical domain that include the temporal variation of the length and circulation scalings as well as the implicit time dependence of the similarity variables. Although the leading/trailing-edge solutions are obtained independent of each other, they are each coupled to the sweeping flow in a simple manner. Specifically, the effect of the incoming/outgoing direction of the free-stream flow at the leading/trailing edges is represented by a change in the sign of the similarity variable in the governing equation. As a combined result, we were able to accurately capture a more complex evolution of the vortex structure, circulation dynamics, and forces exerted on the plate. The approximated flow is acceptably valid for the initial phase of the motion, up to about 2--3 chords of travel. This was corroborated by comparison with corresponding quantities obtained from Navier-Stokes simulations for both impulsive and constant accelerations and at different angles of attack.

An investigation of the separated flow at the leading edge revealed that the vortex sheet there exhibits very high curvature and gives the appearance of a `bent' sheet at the scale of plate chord. It was suggested that a vortex-entrainment sheet, which allows a non-tangential shedding angle, would be more representative of the flow near this irregular edge point where viscous effects such as diffusive entrainment are significant. 

\begin{acknowledgments}
We acknowledge the partial support of the NSF and ONR in this work.
\end{acknowledgments}
\bibliography{RefA2}

\end{document}